\documentclass[twocolumn,superscriptaddress,longbibliography,aps,pra,preprintnumbers]{revtex4-1}

\usepackage{graphicx}
\usepackage{bm}
\usepackage{color}
\usepackage{epstopdf}
\usepackage{amsmath}
\usepackage{amssymb}
\usepackage{epstopdf}

\usepackage[urlcolor=blue,colorlinks=true,citecolor=blue,linkcolor=blue,pdfstartview={FitH},bookmarks=false]{hyperref}

\graphicspath{{fig/}{./fig/}{.}}

\begin{document}

\title{Yu--Shiba--Rusinov states of impurities in a triangular lattice\\ of NbSe$_{2}$ 
       with spin orbit coupling}

\author{Andrzej Ptok}
\email[e-mail: ]{aptok@mmj.pl}
\affiliation{Institute of Nuclear Physics, Polish Academy of Sciences, 
ul. E. Radzikowskiego 152, 31-342 Krak\'{o}w, Poland}
\affiliation{Institute of Physics, M.\ Curie-Sk\l{}odowska University, 
pl. M. Curie-Sk\l{}odowskiej 1, 20-031 Lublin, Poland}

\author{Szczepan G\l{}odzik}
\email[e-mail: ]{szczepan.glodzik@gmail.com}
\affiliation{Institute of Physics, M.\ Curie-Sk\l{}odowska University, 
pl. M. Curie-Sk\l{}odowskiej 1, 20-031 Lublin, Poland}

\author{Tadeusz Doma\'{n}ski}
\email[e-mail: ]{doman@kft.umcs.lublin.pl}
\affiliation{Institute of Physics, M.\ Curie-Sk\l{}odowska University, 
pl. M. Curie-Sk\l{}odowskiej 1, 20-031 Lublin, Poland}

\date{\today}

\begin{abstract}
We study topography of the spin-polarized bound states of magnetic impurities embedded 
in triangular lattice of a superconducting host. Such states have been observed experimentally 
in 2H-NbSe$_{2}$ crystal [G. C. M\'{e}nard {\it et al.}, 
\href{http://dx.doi.org/10.1038/nphys3508}{Nat.Phys. {\bf 11}, 1013 (2015)}] 
and revealed the oscillating particle-hole asymmetry extending to tens of nanometers. 
Using the Bogoliubov--de Gennes approach we explore the Yu--Shiba--Rusinov states
in presence of spin-orbit interaction. We also study the bound states of double 
impurities for several relative positions in a triangular lattice.

\end{abstract}

\maketitle

\section{Introduction}
\label{sec.intro}

Magnetic impurities are detrimental for the Cooper pairs, because they induce 
the spin-polarized subgap states~\cite{balatsky.vekhter.06,koerting.10,heinrich.pascual.17} 
and (when impurities are dense enough) partly suppress or completely fill in the energy gap of  superconducting sample. 
Such in-gap quasiparticles, dubbed Yu--Shiba--Rusinov (YSR) states~\cite{yu.65,shiba.68,rusinov.69}, 
have been observed experimentally in various systems~\cite{yazdani.jones.97,ji.zhang.08,ruby.pientka.15,hatter.heinrich.15,
menard.guissart.15,choi.rubioverdu.17,assouline.feuilletpalma.17,jellinggaard.groverasmussen.16}.
They always exist in pairs, appearing symmetrically with respect to the chemical 
potential (treated here as the `zero-energy' reference level). Their energies can be controlled 
either electrostatically or magnetically~\cite{jellinggaard.groverasmussen.16}. Another 
feature is their  spin-polarization evidenced by the asymmetric conductance 
at opposite voltages~\cite{salkola.balatsky.97,flatte.byers.97}.

The bounds states formed at magnetic impurities in 3-dimensional isotropic superconductors  
are usually characterized by a relatively short spatial extent~\cite{fetter.65}. Contrary to 
that, in 2-dimensional (2D) lattices G.C. M\'{e}nard {\it et al.}~\cite{menard.guissart.15} 
have reported  different behavior, displaying much longer extent of the YSR states
with alternating (oscillating) particle-hole spectral weights. 
Furthermore, the bound states of impurities in superconducting 2H-NbSe$_{2}$~\cite{ugeda.bradley.16} 
with {\it quasi}-2D triangular lattice structure and strong spin-orbit coupling~\cite{xi.wang.16} 
have revealed the long-range coherent structures of a star-shape, whereas molecular dimers 
developed  more complex spatial features~\cite{kezilebieke.dvorak.17}.

\begin{figure}[!b]
\centering
\includegraphics[width=\linewidth]{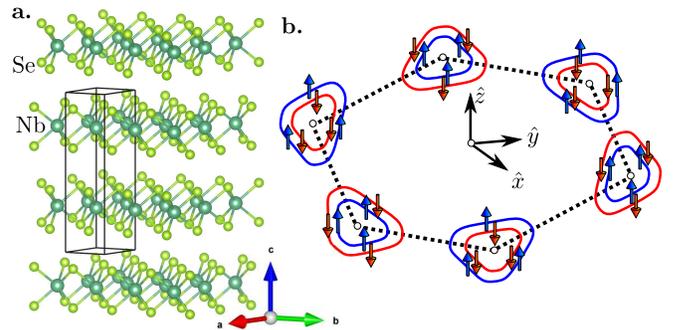}
\caption{
a. Crystallographic structure of the centrosymmetric NbSe$_{2}$ compound and 
its primitive unit cell (black prism). The image was obtained using VESTA 
software~\cite{momma.izumi.11}.
b. Schematic view of the Fermi surface in NbSe$_{2}$ monolayer, adopted from 
Ref.~\cite{bawden.cooil.16}. Blue and red colors correspond to different spin 
(negative/positive) polarizations for each Fermi sheet in the zone-corners.
}
\label{fig.fs}

\end{figure}

Bulk crystals of 2H-NbSe$_{2}$ are characterized by centrosymmetric (P6$_{3}$/mmc) structure, formed by the
stacking of non-centrosymmetric layers~\cite{bawden.cooil.16,meerschaut.deudon.01} (Fig.~\ref{fig.fs}.a).
Every layer is arranged in a honeycomb structure, comprising Nb and Se sublattices.
Local inversion symmetry breaking~\cite{riley.mazzola.14,zhang.liu.14,riley.meevasana.15} 
gives rise to the out-of-plane spin polarization~\cite{bawden.cooil.16} in every layer.
At $T_{CDW} \approx 33 K$ there appears the charge density order~\cite{meerschaut.deudon.01,wilson.disalvo.74,zhu.cao.15}, and below
$T_{c} \approx 7 K$~\cite{bawden.cooil.16} the superconducting state sets in.

The normal state Fermi surface, studied by the angle-resolved photoemission spectroscopy 
(ARPES)~\cite{yokoya.kiss.01,borisenko.kordyuk.09,rahm.hellmann.12,zhu.cao.15,bawden.cooil.16}, 
has revealed two pairs of the Nb-derived pockets, which are  trigonally-warped 
around central $\Gamma$ point and at corner of the (hexagonal) first Brillouin zone.
{\it Ab initio} (DFT) calculations indicated that the Fermi surface sheets originate 
predominantly from Nb 4d-orbitals~\cite{rossnagel.seifarth.01,johannes.mazin.06,flicker.vanwezel.15,bawden.cooil.16}.
In consequence, a triangular lattice formed by Nb atoms plays important role 
for the superconducting state of this compound~\cite{silvaguillen.ordejon.16}  
and implies further a unique star-like shape of the bound states.

Differently than in bulk systems, the Fermi surface of the single monolayer 2H-NbSe$_{2}$ 
consists of  only the pockets around the corner points of the Brillouin zone~\cite{bawden.cooil.16}, 
whose size depends on the spin polarization (Fig.~\ref{fig.fs}.b). The latter effect originates 
from the in plane spin-orbit field~\cite{rahm.hellmann.12,xi.wang.16}. Coupling of the
spin to the {\it valley}  distinguishes between  non-equivalent 
parts of the Brillouin zone. Similar behavior has been also observed in other materials with 
hexagonal lattice structures~\cite{zeng.dai.12,mak.he.12,suzuki.sakano.14,zhu.zeng.14}.

\newpage

Motivated by the recent experimental results of G.C. M\'{e}nard {\it et al.}~\cite{menard.guissart.15}, 
we study  the YSR states of magnetic impurities embedded in a triangular lattice of 
the 2D superconducting host. The single monolayer of the 2H-NbSe$_{2}$ can be treated 
as two dimensional triangular lattice~\cite{silvaguillen.ordejon.16} with in-plane 
spin-orbit field (in the supplementary material we additionally take into account 
also the out-of-plane spin-orbit component which might be realized in other 
compounds~\cite{glass.li.15,sharma.tewari.16}).

In Sec.~\ref{sec.model} we present the microscopic model and discuss some methodological
details. Next, in Sec.~\ref{sec.singleimp}, we analyze the YSR bound states of  
single magnetic impurity in a triangular lattice (Fig.~\ref{fig.schem}), 
focusing on the role of spin-orbit coupling. In Sec.~\ref{sec.doubleimp} we study
the bound states of double magnetic impurities (arranged in 3 different configurations) 
that might be relevant to the experimental data~\cite{kezilebieke.dvorak.17} 
revealing strong  interference effects. Finally  in Sec.~\ref{sec.sum} 
we summarize the main results.

\section{Model and method}
\label{sec.model}

Magnetic impurities embedded in 2D superconducting host with the spin-orbit coupling 
can be described by the following Hamiltonian
\begin{eqnarray}
\label{eq.ham} \hat{\mathcal{H}} = \hat{\mathcal{H}}_{0} + \hat{\mathcal{H}}_{imp} 
+ \hat{\mathcal{H}}_{int} + \hat{\mathcal{H}}_{SOC} .
\end{eqnarray}
The single particle term 
\begin{eqnarray}
\hat{\mathcal{H}}_{0} = - t \sum_{ \langle i,j \rangle \sigma } \hat{c}_{i\sigma}^{\dagger} 
\hat{c}_{j\sigma} - \mu \sum_{i\sigma} \hat{c}_{i\sigma}^{\dagger} \hat{c}_{i\sigma}
\end{eqnarray}
describes the kinetic energy of electrons, where $\hat{c}_{i\sigma}^{\dagger}$ ($\hat{c}_{i\sigma}$) denotes creation (annihilation) of electron with spin $\sigma$ at {\it i}-th lattice site, 
$t$ is a hopping integral between the nearest-neighbors, and $\mu$ is the chemical potential. 
Large spin $S$ of the impurities can be treated classically~
\cite{balatsky.vekhter.06,koerting.10}, and in such case the interaction potential 
can comprise the magnetic $J$ and non-magnetic $K$ parts
\begin{equation}
\hat{\mathcal{H}}_{imp} =  -J \left( \hat{c}_{0\uparrow}^{\dagger} \hat{c}_{0\uparrow}
\!-\! \hat{c}_{0\downarrow}^{\dagger} \hat{c}_{0\downarrow} \right)  
+K \left( \hat{c}_{0\uparrow}^{\dagger} \hat{c}_{0\uparrow}
\!+\! \hat{c}_{0\downarrow}^{\dagger} \hat{c}_{0\downarrow} \right)  .
\end{equation}

We describe the superconducting state, imposing the on-site interaction 
\begin{eqnarray}
\hat{\mathcal{H}}_{int} = U \sum_{i} \hat{c}_{i\uparrow}^{\dagger} \hat{c}_{i\uparrow} 
\hat{c}_{i\downarrow}^{\dagger} \hat{c}_{i\downarrow}
\end{eqnarray}
with attractive potential $U < 0$. Such effective pairing is assumed to be weak, 
therefore we can treat it within the standard mean-field decoupling
\begin{eqnarray}
\hat{c}_{i\uparrow}^{\dagger} \hat{c}_{i\uparrow} \hat{c}_{i\downarrow}^{\dagger} 
\hat{c}_{i\downarrow} &=& \chi_{i} \hat{c}_{i\uparrow}^{\dagger} \hat{c}_{i\downarrow}^{\dagger} 
+ \chi_{i}^{\ast} \hat{c}_{i\downarrow} \hat{c}_{i\uparrow} - | \chi_{i} |^{2} \\
\nonumber &+& n_{i\uparrow} \hat{c}_{i\downarrow}^{\dagger} \hat{c}_{i\downarrow} 
+ n_{i\downarrow} \hat{c}_{i\uparrow}^{\dagger} \hat{c}_{i\uparrow} 
- n_{i\uparrow} n_{i\downarrow} ,
\end{eqnarray}
where $\chi_{i} = \langle \hat{c}_{i\downarrow} \hat{c}_{i\uparrow} \rangle$ is the superconducting order parameter and $n_{i\sigma} = \langle \hat{c}_{i\sigma}^{\dagger} \hat{c}_{i\sigma} \rangle$ is the average number of spin $\sigma$ particles  at {\it i}-th site. 
Hartree term can be incorporated into the {\it effective} local spin-dependent chemical 
potential $\mu \rightarrow \tilde{\mu}_{i\sigma} \equiv \mu - U n_{i\bar{\sigma}}$.
As we shall see, impurities suppress the local order parameter $\chi_{i}$  whose magnitude 
and sign depend on the coupling strength $J$~\cite{smith.tanaka.16,goertzen.tanaka.17}.

\begin{figure}[!t]
\centering
\includegraphics[width=0.5\linewidth]{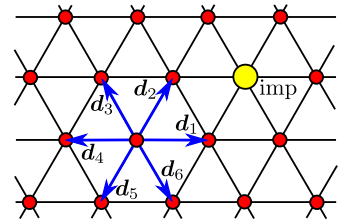}
\caption{
Schematic illustration of a single magnetic impurity (yellow) in a 2-dimensional 
triangular lattice. Each lattice site is surrounded by six nearest neighbors 
at positions ${\bm d}_{\alpha}$.
\label{fig.schem}
}
\end{figure}

The spin-orbit coupling (SOC) can be expressed by~\cite{xu.qu.14} 
\begin{eqnarray}
\hat{\mathcal{H}}_{SOC} = - i \lambda \sum_{ij\sigma\sigma'} \hat{c}_{i+\bm{d}_{j}\sigma}^{\dagger} \left( \bm{d}_{j} \times 
\hat{\bm \sigma}^{\sigma\sigma'} \right) \cdot \hat{w} \; \hat{c}_{i\sigma'} ,
\end{eqnarray}
where vector ${\bm d_j} = ( d_{j}^{x} ,  d_{j}^{y} , 0 )$ corresponds to the nearest neighbors 
of {\it i}-th site (Fig.~\ref{fig.schem}), and $\hat{\bm \sigma} = ( \sigma_{x} , \sigma_{y} , \sigma_{z} )$ consists of the Pauli matrices.
The unit vector $\hat{w}$ shows a direction of the spin orbit field, which in general can be arbitrary, 
but  we restrict our considerations to the in-plane $\hat{w} \equiv \hat{x} = ( 1 , 0 , 0 )$ and out-of-plane $\hat{z} = ( 0 , 0 , 1 )$ polarizations,  so formally we have
\begin{eqnarray}
\nonumber \left( \bm{d}_{j} \times \hat{\bm \sigma} \right) \cdot \hat{w} = 
\left\lbrace
\begin{array}{ll}
d_{j}^{y} \sigma_{z} & \mbox{for in-plane field,} \\ 
d_{j}^{x} \sigma_{y} - d_{j}^{y} \sigma_{x} & \mbox{for out-of-plane field.}
\end{array} 
\right. \\
\end{eqnarray}
Let us notice, that out-of-plane component mixes $\uparrow$ and $\downarrow$ particles, whereas the in plane field corresponds to additional spin-conserving hopping with the spin- and direction-dependent amplitude.

\subsection*{Bogoliubov--de Gennes technique}

Magnetic impurities break the translational invariance of the system, therefore the local pairing amplitude $\chi_{i}$ and occupancy $n_{i\sigma}$ have to be determined for each lattice site individually~\cite{ptok.kapcia.15}. 
One can diagonalize the Hamiltonian~(\ref{eq.ham})  via the following unitary transformation
\begin{eqnarray}
\label{eq.bvtransform} c_{i\sigma} = \sum_{n} \left( u_{in\sigma} \gamma_{n} 
- \sigma v_{in\sigma}^{\ast} \gamma_{n}^{\dagger} \right) 
\end{eqnarray}
where $\gamma_{n}$ and  $\gamma_{n}^{\dagger}$ are  {\it quasi}-particle fermionic operators, with the eigenvectors $u_{in\sigma}$ and $v_{in\sigma}$. 
This leads to the Bogoliubov--de~Gennes (BdG) equations~\cite{degennes.99}
\begin{eqnarray}
\label{eq.bdg} \mathcal{E}_{n} &&
%%%%%%
\left(
\begin{array}{c}
u_{in\uparrow} \\ 
v_{in\downarrow} \\ 
u_{in\downarrow} \\ 
v_{in\uparrow}
\end{array} 
\right) \\
%%%%%%
\nonumber = \sum_{j} && \left(
\begin{array}{cccc}
H_{ij\uparrow} & D_{ij} & S_{ij}^{\uparrow\downarrow} & 0 \\ 
D_{ij}^{\ast} & -H_{ij\downarrow}^{\ast} & 0 & S_{ij}^{\downarrow\uparrow} \\ 
S_{ij}^{\downarrow\uparrow} & 0 & H_{ij\downarrow} & D_{ij} \\ 
0 & S_{ij}^{\uparrow\downarrow} & D_{ij}^{\ast} & -H_{ij\uparrow}^{\ast}
\end{array} 
\right) 
%%%%%%
\left(
\begin{array}{c}
u_{jn\uparrow} \\ 
v_{jn\downarrow} \\ 
u_{jn\downarrow} \\ 
v_{jn\uparrow}
\end{array} 
\right)
\end{eqnarray}
containing the single-particle term 
$H_{ij\sigma} = - t \delta_{\langle i,j \rangle} 
- \left( \tilde{\mu}_{i\sigma} + ( K - \sigma J ) \delta_{i0} \right) \delta_{ij} 
+ S_{ij}^{\sigma\sigma}$ 
and the spin-orbit coupling term
$S_{ij}^{\sigma\sigma'} = - i \lambda  \sum_{l} \left( {\bm d}_{l} \times 
\hat{\bm \sigma}^{\sigma\sigma'} \right) \cdot \hat{w} \; \delta_{j,i+{\bm d}_{l}}$.
Here, $S_{ij}^{\sigma\sigma}$ and $S_{ij}^{\sigma\bar{\sigma}}$ correspond to in plane and out of plane spin orbit field, respectively, which satisfy $S_{ij}^{\sigma\sigma'} = ( S_{ji}^{\sigma'\sigma} )^{\ast}$ and
$D_{ij} = U \chi_{i} \delta_{ij}$ describes the on-site pairing.
The superconducting order parameter $\chi_{i}$ and occupancy $n_{i\sigma}$ can be computed self-consistently from BdG equations~(\ref{eq.bdg})
\begin{eqnarray}
\chi_{i} &=&  
\sum_{n} \left[ u_{in\downarrow} v_{in\uparrow}^{\ast} f( \mathcal{E}_{n} ) 
- u_{in\uparrow} v_{in\downarrow}^{\ast} f ( - \mathcal{E}_{n} ) \right] , \\
n_{i\sigma} &=&  
\sum_{n} \left[ | u_{in\sigma} |^{2} f( \mathcal{E}_{n} ) + 
| v_{in\bar{\sigma}} |^{2} f ( - \mathcal{E}_{n} ) \right] ,
\end{eqnarray}
where $ f ( \omega ) = 1 / \left[ 1 + \exp ( \omega / k_{B} T ) \right]$ 
is the Fermi-Dirac distribution.
In particular, the spin-resolved local density of states (LDOS) is
given by~\cite{matsui.sato.03}
\begin{eqnarray}
\nonumber \rho_{i\sigma} ( \omega ) = \sum_{n} \left[ | u_{in\sigma} |^{2} \delta 
( \omega - \mathcal{E}_{n} ) + | v_{in\sigma} |^{2} \delta ( \omega + \mathcal{E}_{n} ) 
\right] . \\
\label{eq.ldos}
\end{eqnarray}
For its numerical determination we have replaced the Dirac 
delta function by Lorentzian $\delta (\omega) = \zeta / [ \pi ( \omega^{2} 
+ \zeta^{2})]$ with a small broadening $\zeta =0.025 t$.

\section{Numerical results and discussion}
\label{sec.num}

We now present the BdG results obtained for the single impurity embedded in 
a triangular lattice (Sec.~\ref{sec.singleimp}) 
and for several configurations of two magnetic impurities (Sec.~\ref{sec.doubleimp}).
Numerical computations have been done at zero temperature for the finite cluster 
$N_{a} \times N_{b} = 41 \times 41$, assuming $U/t = - 3$, $\mu/t = 0$, $K/t = 0$, 
and determining the bound states for varying $J$. In this work we focus on the effect of in-plane spin-orbit field, and  additional results for the out-of-plane SOC are shown in the Supplemental Material (SM)~\cite{sm.pdf}.

\begin{figure}[!t]
\centering
\includegraphics[width=\linewidth]{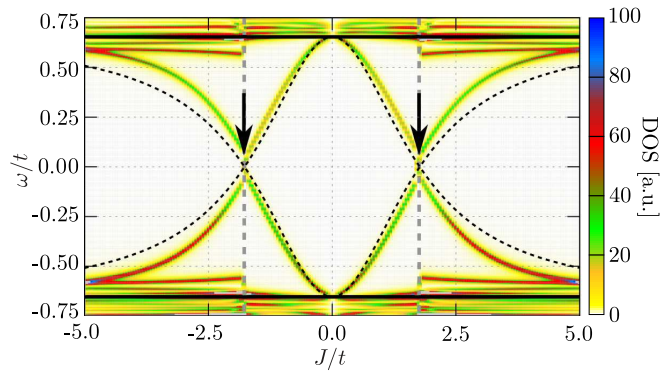}
\caption{Evolution of the low energy spectrum with respect to the magnetic coupling $J$.
Solid black line shows magnitude of the pairing gap in regions far away from the impurity, 
black arrows 
point at the quantum phase transition (i.e.\ crossing of the subgap YSR states), and 
thin-dashed lines display the  YSR bound states  calculated 
from Eq.~(\ref{eq.eysr}). Results obtained without the SOC.
\label{fig.one_dos}
}
\end{figure}

\subsection{Single magnetic impurity}
\label{sec.singleimp}

Let us start by discussing the  results obtained in absence of the spin-orbit 
coupling. Typical quasiparticle spectrum is displayed in Fig.~\ref{fig.one_dos},
where we can recognize the gaped region $|\omega |\leq \Delta$ of superconducting 
host (for our set of the model parameters $\Delta \simeq 0.65 t$) and one 
pair of the bound states, appearing symmetrically around the chemical potential. 
Energies $\pm E_{\alpha}$ of these  states and spectral weights depend on the 
coupling $J$. In particular, at some critical $J_{c}$ (indicated by black arrows) 
they eventually cross each other. This crossing is a hallmark of the quantum 
phase transition (QPT)~\cite{sakurai.70} in which the ground state undergoes 
qualitative evolution~\cite{salkola.balatsky.97}. When magnetic coupling  overcomes 
the pairing energy (for $|J| \geq J_{c}$) the particle and hole states become degenerate, 
and the ground state changes from a BCS singlet to a spinful configuration~\cite{salkola.balatsky.97,
vangervenoei.tanaskovic.17,franke.schulze.11,kaladzhyan.bena.16}.

\begin{figure}[!b]
\centering
\includegraphics[width=\linewidth]{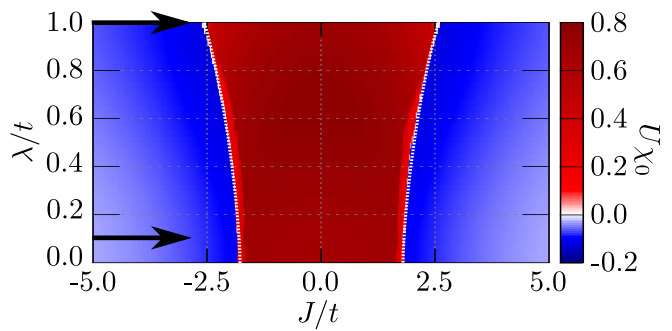}
\caption{
Influence of the in-plane SOC on the critical value $J_{c}$. 
Blue/red colors correspond to the discontinuous change of $U \chi_{0}$ at 
the impurity site, and a white line marks the QPT. Black arrows indicate
two values of $\lambda$, for which the profiles are shown in Fig.~\ref{fig.qcp}.
\label{fig.df_in}
}
\end{figure}

\begin{figure}[!t]
\centering
\includegraphics[width=\linewidth]{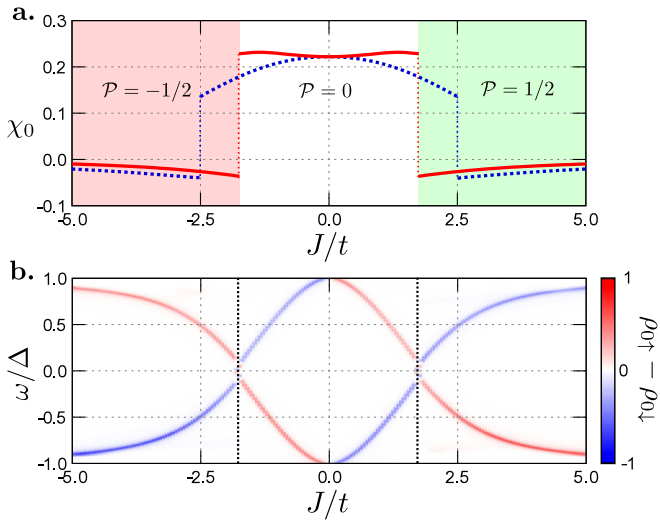}
\caption{
The order parameter $\chi_{0}=\langle c_{0\downarrow} c_{0\uparrow} \rangle$  obtained 
at the impurity site $i=0$ (panel a) for the weak (red line) and strong (blue dotted line) 
spin-orbit couplings, with $\lambda/t=0.1$ and $1.0$ respectively. Magnetic polarization 
of the YSR states $\rho_{0\uparrow}(\omega) - \rho_{0\downarrow}(\omega)$ (panel b), 
obtained for $\lambda/t = 0.1$.
\label{fig.qcp}
}
\end{figure}

Our BdG data can be confronted with the analytical results of the thermodynamic
limit $N_{a} \times N_{b} \rightarrow \infty$~\cite{vangervenoei.tanaskovic.17}:
\begin{eqnarray}
E_{\mbox{YSR}} = \pm \Delta \frac{1 - \alpha^{2}}{1 + \alpha^{2}} ,
\label{eq.eysr} 
\end{eqnarray}
where $\alpha = \pi \rho_{0} J$ is the dimensionless impurity coupling 
parameter, $\rho_{0}$ is the normal state DOS at the Fermi level, and $\Delta$ 
is the superconducting gap. These quasiparticle energies~(\ref{eq.eysr}) are
displayed in Fig.~\ref{fig.one_dos} by a thin-dashed line. In the weak coupling limit 
$|J| \leq J_{c}$, the formula~(\ref{eq.eysr}) matches well with our numerical 
BdG results. Some differences appear above the QPT (for $|J| \geq J_{c}$), where 
the local pairing parameter at magnetic impurity is substantially reduced affecting also the pairing gap of its neighboring sites. With an increasing coupling $\lambda$, the QPT is shifted to higher values (Fig.~\ref{fig.df_in}). The critical $J_{c}$ corresponds to value of $J$ at which the YSR states cross each other. Variation of the critical $J_{c}$ is caused by influence of the SOC merely on the normal state DOS ($\rho_{0}$).

Such quantum phase transition is manifested by a sign change of the order parameter $\chi_0$ 
at the impurity site (Fig.~\ref{fig.qcp}.a) and discontinuity of its absolute value is 
a signature of the first-order phase transition~\cite{mashkoori.bjornson.17,pershoguba.bjornson.15,glodzik.ptok.17,mashkoori.bjornson.17}.
Let us emphasize, that QPT is associated also with a reversal of the YSR polarization 
(Fig.~\ref{fig.qcp}.b) and furthermore the total polarization of the system 
$\mathcal{P} = \frac{1}{2} \sum_{i} \left( n_{i\uparrow} - n_{i\downarrow} 
\right)$ abruptly changes at $J= \pm J_{c}$ from zero to $\pm 1/2$~\cite{morr.stavropoulos.03}. 
Similar behavior can be observed for multiple impurities~\cite{morr.yoon.06}.

\begin{figure}[!t]
\centering
\includegraphics[width=\linewidth]{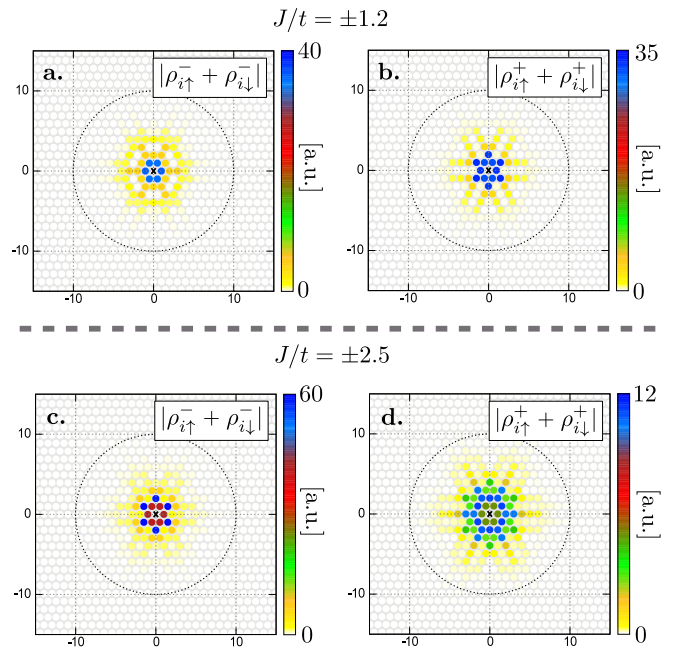}
\caption{
Spatial patterns of the ``negative'' and ``positive'' YSR states
$| \bar{\rho}^{\pm}_{i\uparrow} + \bar{\rho}^{\pm}_{i\downarrow}|$.
Results are obtained for $J/t = \pm 1.2$ (top panel) and $J/t = \pm 2.5$
(bottom panel), assuming the in-plane spin orbit coupling $\lambda/t=0.1$.
\label{fig.one_ldos}
}
\end{figure}

\begin{figure}[!b]
\centering
\includegraphics[width=\linewidth]{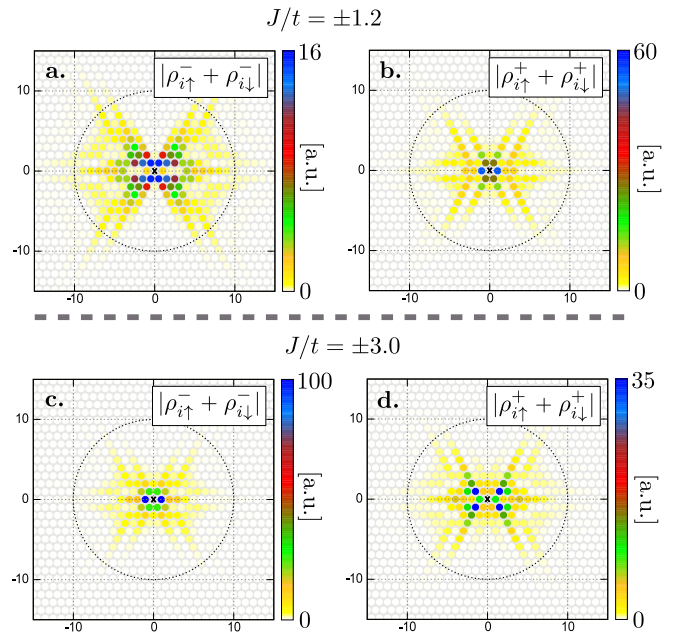}
\caption{
The same as in Fig.~\ref{fig.one_ldos}, but for the stronger spin-orbit 
interaction $\lambda/t = 1$.
\label{fig.bigsoc_ldos}
}
\end{figure}

In the weak coupling limit (i.e.\ for $\lambda \ll t$) we can hardly notice any 
meaningful influence of SOC on the bulk superconductivity and the YSR states (see 
Fig.~1 in the SM~\cite{sm.pdf}). Similar conclusion has been previously 
reported from the $T$-matrix treatment of magnetic impurities for 1D and 2D square
lattices by V.\ Kaladzhyan {\it et al.}~\cite{kaladzhyan.bena.16}. Our calculations
have been done for $\lambda/t = 0.1$ which could to be realistic for NbSe$_{2}$ compound. 
Obviously for much stronger values of the spin-orbit coupling, both the superconducting
state and the bound YSR states depend on magnitude of $\lambda$ and direction 
of the magnetic moment~\cite{kim.zhang.15}.

Let us now explore a spatial extent of the YSR states. This can be achieved 
within the BdG approach by integrating the spectral weights 
\begin{eqnarray}
\rho_{i\sigma}^{\pm} = \int_{\omega_{1}}^{\omega_{2}} \rho_{i\sigma}(\omega) \; d\omega
\end{eqnarray}
in the interval $\omega \in (\omega_{1}, \omega_{2})$ capturing every in-gap  quasiparticle 
below or above the Fermi level~\cite{rontynen.ojanen.15}.
Our numerical calculations have been done for the single impurity in the weak 
$J = -1.2 t > J_{c}$ and strong magnetic coupling limits $J = -2.5t < J_{c}$, 
respectively. The results shown in 
Fig.~\ref{fig.one_ldos} (notice different scales for each of these panels) reveal
the characteristic  6-leg star shape, whose extent spreads on several sites 
around the magnetic impurity. Spectral weights at the positive and negative 
energies are quite different, leading to a finite spin polarization of the YSR 
states (displayed in the bottom panel in Fig.~\ref{fig.qcp}).

\begin{figure}[!b]
\centering
\includegraphics[width=\linewidth]{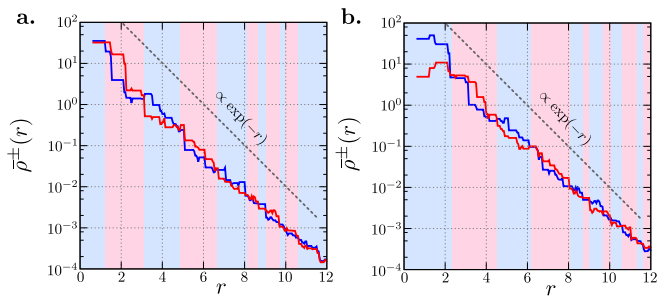}
\caption{
Profiles of the hole- (blue line) and electron-like (red line) displaced moving 
average (DMA) for the YSR bound states $\bar{\rho}^{\pm} ( r )$ as a function 
of distance $r$ from the impurity (with $\delta r = 0.5$). The left and right 
panels correspond to  $| J | < J_{c}$ and  $| J | > J_{c}$, respectively.
Results are obtained for the in-plane spin orbit field $\lambda / t = 0.1$.
The dashed gray line corresponds to  $\exp ( - r )$, which is a guide to eye.
The blue/red background color indicates the dominant hole/particle type of 
YSR state at a given $r$.
\label{fig.one_dma}
}
\end{figure}

Upon varying $J$ we observe, that the star-shape (characterizing $C_{6}$ symmetry 
of a triangular lattice) is rather robust. Such patterns of YSR states could be probed 
by the scanning tunneling microscopy, which nowadays has an atomic scale resolution~\cite{ji.zhang.08,randeria.feldman.16,choi.rubioverdu.17,kim.yoshida.17,reecht.heinrich.17}. 
Let us emphasize that this 6-leg star shape originates from a triangular lattice geometry 
and from particular topology of the Fermi surface~\cite{weismann.wenderoth.09} in agreement 
with the experimental observations~\cite{menard.guissart.15}. We have checked that the YSR 
states are only quantitatively (by a few percent) affected by the weak ($\lambda / t=0.1$) 
in-plane spin-orbit coupling. For stronger SOC the results are presented in 
Fig.~\ref{fig.bigsoc_ldos}. Comparison with Fig.~\ref{fig.one_ldos} shows that for 
higher $\lambda$ the YSR states gradually increase their extent, and the star-like 
shape seems to be weakly deformed with some elongation parallel to $x$-axis. 
Fig.~\ref{fig.bigsoc_ldos}.a presents especially large range of bound states, 
extending well beyond ten lattice constants from the impurity site. As stated in 
the previous section, in-plane spin-orbit field leads to the presence of 
$S^{\sigma\sigma}_{ij}$ term in the single particle part of BdG equations, which 
directly affects the hopping amplitude. To observe influence of this term on 
the spectral function, $\lambda$ has to be of the order of $t$. We suspect, 
that large spatial extent of the YSR states reported in~\cite{menard.guissart.15} 
was a consequence of the reduced dimensionality and/or the structure of atomic 
lattice, and was rather not much affected by the in-plane spin orbit field of 
$NbSe_{2}$. In general, however, materials with the stronger SOC couplings could 
reveal some increase in the spatial extent of subgap bound states.

For some quantitative analysis of the spatial profiles of YSR states we define 
the {\em displaced moving average} (DMA) $\bar{\rho}^{\pm} ( r )$ interpreted 
as an averaged spectral weight contained in a ring of radius $r$ and its 
half-width $\delta r$. It depends only on a radial distance $r$ from the magnetic 
impurity ${\bm r}_{0}$, averaging the angle-dependent fluctuations. Our results 
are presented in Fig.~\ref{fig.one_dma}. They clearly show, that functions 
$\bar{\rho}^{\pm}(r)$ of the YSR states are characterized by  particle and 
hole oscillations that are opposite in phase (see the blue and red lines). 
Such particle-hole oscillations decay exponentially with  distance  
(notice a logarithmic scale in Fig.~\ref{fig.one_dma} in agreement with 
previous studies~\cite{morr.stavropoulos.03,kawakami.hu.15,menard.guissart.15}. 
In 2D continuum version of this model the wavefunctions of the YSR states
have been expressed analytically~\cite{menard.guissart.15}:
\begin{eqnarray}
\psi^{\pm} ( r ) \propto  \frac{1}{ \sqrt{ k_{F} r } } \sin \left( k_{F} r - \frac{ \pi }{ 4 } + \delta^{\pm} \right) \\
\nonumber \times \exp \left[ - \sin ( \delta^{+} - \delta^{-} ) \frac{ r }{ \zeta } \right],
\end{eqnarray}
where $k_{F}$ is the Fermi wave vector, $r$ is the distance from the impurity, 
whereas $\zeta$ is the superconducting coherence length. Both functions oscillate 
with $k_{F}r$, but with different scattering phase shifts $\delta^{\pm} = K \rho_{0} 
\pm J / \rho_{0}$. At short distances the YSR wavefunctions are governed by 
$\sin ( k_{F} r ) / \sqrt{ k_{F} r }$, whereas for larger $r$ the exponential envelope 
function suppresses particle-hole oscillations (dotted line in Fig.~\ref{fig.one_dma}). 
Dominant (particle or hole) contributions to the YSR bound states are displayed
by an alternating color of the background in Fig.~\ref{fig.one_dma}. The period 
of such oscillations is approximately equal to $\sim 2$ lattice constants. 
Out-of-plane spin orbit field leads to a similar behavior (Fig.~5 in~\cite{sm.pdf}).

\begin{figure}[!t]
\centering
\includegraphics[width=\linewidth]{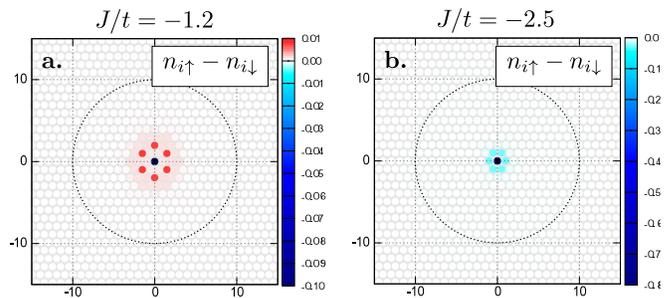}
\caption{
Magnetization along $z$-axis induced near the magnetic impurity 
for $|J|<J_{c}$ (panel a) and  $|J|>J_{c}$ (panel b).
\label{fig.magnetization}
}
\end{figure}

Quantum phase transition (at $J_{c}$) has consequences on a reversal of the 
magnetization induced near the impurity (see Fig.~\ref{fig.magnetization}).
For $| J | < J_{c}$ the impurity is weakly screened, whereas for stronger couplings
$| J | > J_{c}$ the impurity polarizes its neighborhood in the direction of its 
own magnetic moment. In both cases, this short-range magnetization does not 
coincide with the six-leg-star shape of the bound states. Differences between 
the YSR wave-functions and various components of magnetization have been previously 
discussed for 2D square lattice by V.~Kaladzhyan {\em et al.}~\cite{kaladzhyan.bena.16}. 

%%%%%%%%%%%%%%%%%%%%%%%%%%%%%%%
%%%%%%%%%%%%%%%%%%%%%%%%%%%%%%%
%%%%%%%%%%%%%%%%%%%%%%%%%%%%%%%

\subsection{YSR of double impurities}
\label{sec.doubleimp}

\begin{figure}
\centering
\includegraphics[width=\linewidth]{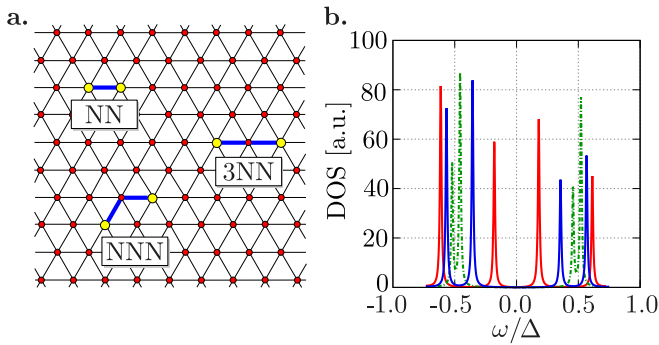}
\caption{
Schematic illustration of two magnetic impurities arranged in three different 
configurations (a) and the subgap spectrum (b) for the nearest neighbors 
(NN), next nearest neighbors (NNN) and the third nearest neighbors (3NN) as shown 
by red, blue and green lines, respectively. We assumed the in-plane spin orbit 
coupling $\lambda/t = 0.1$.
\label{fig.two_dos}
}
\end{figure}

BdG technique has a virtue that it can be easily applied for studying the bound 
states of more numerous impurities, distributed at arbitrary positions in 
a crystallographic lattice. In this section we consider the case of double 
magnetic impurities arranged in three different configurations displayed 
in Fig.~\ref{fig.two_dos}.a. Our study of the YSR states is inspired by 
the results of Ref.~\cite{kezilebieke.dvorak.17} for ferromagnetic dimers. 
Such BdG  calculations can be applied to more complex molecules~\cite{franke.schulze.11,jacob.soriano.13,kugel.karolak.15,island.gaudenzi.17} and/or 
multi-impurity structures~\cite{nakosai.tanaka.13,menard.guissart.16,lado.fernandezrossier.16}.
It is well known~\cite{balatsky.vekhter.06}, that multiple-impurities can develop
several quantum phase transitions with some characteristic features. They have 
been previously studied for 2D lattices, treating the spins classically~
\cite{morr.yoon.06,hoffman.klinovaja.15,meng.klinovaja.15} and taking into 
account the strong correlation effects within the Anderson-type scenario~
\cite{zitko.15}. Here we  explore the YSR states of two classical magnetic 
impurities embedded in a triangular lattice, assuming the weak in-plane 
spin orbit interaction $\lambda/t = 0.1$.

\begin{figure}[!t]
\centering
\includegraphics[width=\linewidth]{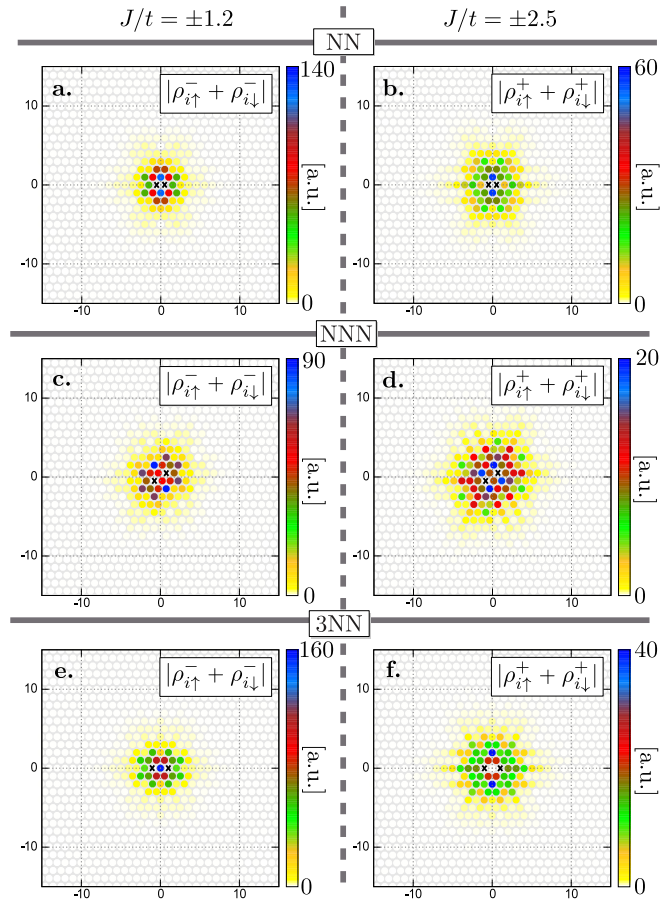}
\caption{Spatial pattern of the YSR states for the double magnetic impurities in different 
(NN, NNN, 3NN) configurations, as indicated. The left column panels refer to the weak 
coupling $J / t = - 1.2$ and the right column panels to the strong coupling $J / t = -2.5$
limits, respectively. We imposed the in-plane spin orbit coupling $\lambda / t = 0.1$
that could be relevant to NbSe$_{2}$.
\label{fig.two_soc}}
\end{figure}

Fig.~\ref{fig.two_dos} presents the subgap spectrum obtained for different configurations 
of the double impurities. We notice that coupling between the impurities induces the double-peak 
structure of YSR states, both at negative and positive energies (panel b). 
Fig.~\ref{fig.two_soc} displays spatial distributions of the YSR states for each configuration 
of the double magnetic impurities for the weak (left column) and strong (right column) couplings 
$J$. Although the $C_6$ rotational symmetry is broken, one can clearly see that mirror-symmetry 
with respect to the axis connecting these double-impurities and the axis perpendicular to it.
Novel spatial patterns of the YSR states are due to the constructive/destructive quantum 
interference between the overlapping subgap states. Obviously, the most significant 
quantum interference occurs for the quantum impurities either at the nearest neighbor 
(NN) or next nearest neighbor (NNN)  configurations, with clear a bonding-antibonding
splitting of the YSR quasiparticle energies. For more distant arrangements of the double
impurities (for instance 3NN) their spatial patterns gradually evolve back to the star-like 
shape. A more in-depth comparison of the results obtained with and without SOC is presented 
in the SM~\cite{sm.pdf} Fig.~6 and 7. We hope that our theoretical predictions could be 
empirically verified, using the combined AFM (capable of manipulating the impurities) 
and STM (suitable for probing the subgap spectrum) techniques.

\section{Conclusions}
\label{sec.sum}

In summary, we have investigated the energies and spatial extent of the Yu--Shiba--Rusinov 
(YSR) states induced by the classical magnetic impurities embedded in a 2-dimensional triangular 
lattice of 2H-NbSe$_{2}$ superconducting host. To study this particular crystallographic geometry 
in presence of local inhomogeneities (in a form of the single or double impurities) and 
the spin-orbit coupling (SOC) we have adopted the Bogoliubov--de Gennes formalism. 

In agreement 
with the experimental observations~\cite{menard.guissart.15} we have found that the YSR states 
acquire the 6-fold rotational symmetry (star-like patterns) whose spectroscopic signatures 
extend onto about a dozen of intersite distances. Furthermore, their intertwining ($\pi$-shifted 
in phase) particle-hole oscillations are clearly visible. The weak spin orbit coupling (which 
should be relevant to 2H-NbSe$_{2}$ compound) has rather negligible influence on the energies 
of YSR states, but for relatively stronger SOC their spatial extent eventually increases 
(beyond 10 lattice constants in some cases). Analysis of the SOC for the single impurity
indicates, that the extended range of YSR states reported in~\cite{menard.guissart.15} stems 
from the dimensionality and/or the structure of atomic lattice, rather than from the in-plane 
spin-orbit field of such materials. 

We have also studied the subgap quasiparticle spectrum of double magnetic impurities in 
three different configurations, revealing either the constructive or destructive quantum 
interference which breaks $C_{6}$ symmetry of the YSR wave-functions. Deviation from 
the star-like shape (typical for single impurities)  depends on the relative distance 
between such magnetic impurities. When they are close to each other, the YSR states 
develop a double-peak structure (characteristic for the bonding and antibonding states) 
whose spatial patters no longer resemble the star shape. With an increasing distance 
between the impurities such bonding-antibonding splitting gradually disappears, and 
the spatial star-like shape of YSR states is gradually restored.

\begin{acknowledgments}
We thank K.J. Kapcia for valuable comments and discussions. This work was supported by the National Science Centre (NCN, Poland) under grants UMO-2016/20/S/ST3/00274 (A.P.) and DEC-2014/13/B/ST3/04451 (S.G. and T.D.)
\end{acknowledgments}

\bibliography{biblio}

\begin{thebibliography}{73}%
\makeatletter
\providecommand \@ifxundefined [1]{%
 \@ifx{#1\undefined}
}%
\providecommand \@ifnum [1]{%
 \ifnum #1\expandafter \@firstoftwo
 \else \expandafter \@secondoftwo
 \fi
}%
\providecommand \@ifx [1]{%
 \ifx #1\expandafter \@firstoftwo
 \else \expandafter \@secondoftwo
 \fi
}%
\providecommand \natexlab [1]{#1}%
\providecommand \enquote  [1]{``#1''}%
\providecommand \bibnamefont  [1]{#1}%
\providecommand \bibfnamefont [1]{#1}%
\providecommand \citenamefont [1]{#1}%
\providecommand \href@noop [0]{\@secondoftwo}%
\providecommand \href [0]{\begingroup \@sanitize@url \@href}%
\providecommand \@href[1]{\@@startlink{#1}\@@href}%
\providecommand \@@href[1]{\endgroup#1\@@endlink}%
\providecommand \@sanitize@url [0]{\catcode `\\12\catcode `\$12\catcode
  `\&12\catcode `\#12\catcode `\^12\catcode `\_12\catcode `\%12\relax}%
\providecommand \@@startlink[1]{}%
\providecommand \@@endlink[0]{}%
\providecommand \url  [0]{\begingroup\@sanitize@url \@url }%
\providecommand \@url [1]{\endgroup\@href {#1}{\urlprefix }}%
\providecommand \urlprefix  [0]{URL }%
\providecommand \Eprint [0]{\href }%
\providecommand \doibase [0]{http://dx.doi.org/}%
\providecommand \selectlanguage [0]{\@gobble}%
\providecommand \bibinfo  [0]{\@secondoftwo}%
\providecommand \bibfield  [0]{\@secondoftwo}%
\providecommand \translation [1]{[#1]}%
\providecommand \BibitemOpen [0]{}%
\providecommand \bibitemStop [0]{}%
\providecommand \bibitemNoStop [0]{.\EOS\space}%
\providecommand \EOS [0]{\spacefactor3000\relax}%
\providecommand \BibitemShut  [1]{\csname bibitem#1\endcsname}%
\let\auto@bib@innerbib\@empty
%</preamble>
\bibitem [{\citenamefont {Balatsky}\ \emph {et~al.}(2006)\citenamefont
  {Balatsky}, \citenamefont {Vekhter},\ and\ \citenamefont
  {Zhu}}]{balatsky.vekhter.06}%
  \BibitemOpen
  \bibfield  {author} {\bibinfo {author} {\bibfnamefont {A.~V.}\ \bibnamefont
  {Balatsky}}, \bibinfo {author} {\bibfnamefont {I.}~\bibnamefont {Vekhter}}, \
  and\ \bibinfo {author} {\bibfnamefont {J.-X.}\ \bibnamefont {Zhu}},\
  }\bibfield  {title} {\enquote {\bibinfo {title} {Impurity-induced states in
  conventional and unconventional superconductors},}\ }\href {\doibase
  10.1103/RevModPhys.78.373} {\bibfield  {journal} {\bibinfo  {journal} {Rev.
  Mod. Phys.}\ }\textbf {\bibinfo {volume} {78}},\ \bibinfo {pages} {373}
  (\bibinfo {year} {2006})}\BibitemShut {NoStop}%
\bibitem [{\citenamefont {Koerting}\ \emph {et~al.}(2010)\citenamefont
  {Koerting}, \citenamefont {Andersen}, \citenamefont {Flensberg},\ and\
  \citenamefont {Paaske}}]{koerting.10}%
  \BibitemOpen
  \bibfield  {author} {\bibinfo {author} {\bibfnamefont {V.}~\bibnamefont
  {Koerting}}, \bibinfo {author} {\bibfnamefont {B.~M.}\ \bibnamefont
  {Andersen}}, \bibinfo {author} {\bibfnamefont {K.}~\bibnamefont {Flensberg}},
  \ and\ \bibinfo {author} {\bibfnamefont {J.}~\bibnamefont {Paaske}},\
  }\bibfield  {title} {\enquote {\bibinfo {title} {Nonequilibrium transport via
  spin-induced subgap states in superconductor/quantum dot/normal metal
  cotunnel junctions},}\ }\href {\doibase 10.1103/PhysRevB.82.245108}
  {\bibfield  {journal} {\bibinfo  {journal} {Phys. Rev. B}\ }\textbf {\bibinfo
  {volume} {82}},\ \bibinfo {pages} {245108} (\bibinfo {year}
  {2010})}\BibitemShut {NoStop}%
\bibitem [{\citenamefont {Heinrich}\ \emph {et~al.}(2017)\citenamefont
  {Heinrich}, \citenamefont {Pascual},\ and\ \citenamefont
  {Franke}}]{heinrich.pascual.17}%
  \BibitemOpen
  \bibfield  {author} {\bibinfo {author} {\bibfnamefont {B.~W.}\ \bibnamefont
  {Heinrich}}, \bibinfo {author} {\bibfnamefont {J.~I.}\ \bibnamefont
  {Pascual}}, \ and\ \bibinfo {author} {\bibfnamefont {K.~J.}\ \bibnamefont
  {Franke}},\ }\href@noop {} {\enquote {\bibinfo {title} {Single magnetic
  adsorbates on s-wave superconductors},}\ } (\bibinfo {year} {2017}),\ \Eprint
  {http://arxiv.org/abs/arXiv:1705.03672} {arXiv:1705.03672} \BibitemShut
  {NoStop}%
\bibitem [{\citenamefont {Yu}(1965)}]{yu.65}%
  \BibitemOpen
  \bibfield  {author} {\bibinfo {author} {\bibfnamefont {L.}~\bibnamefont
  {Yu}},\ }\bibfield  {title} {\enquote {\bibinfo {title} {Bound state in
  superconductors with paramagnetic impurities},}\ }\href {\doibase
  10.7498/aps.21.75} {\bibfield  {journal} {\bibinfo  {journal} {Acta Phys.
  Sin}\ }\textbf {\bibinfo {volume} {21}},\ \bibinfo {pages} {75} (\bibinfo
  {year} {1965})}\BibitemShut {NoStop}%
\bibitem [{\citenamefont {Shiba}(1968)}]{shiba.68}%
  \BibitemOpen
  \bibfield  {author} {\bibinfo {author} {\bibfnamefont {H.}~\bibnamefont
  {Shiba}},\ }\bibfield  {title} {\enquote {\bibinfo {title} {Classical spins
  in superconductors},}\ }\href {\doibase 10.1143/PTP.40.435} {\bibfield
  {journal} {\bibinfo  {journal} {Progr. Theor. Exp. Phys.}\ }\textbf {\bibinfo
  {volume} {40}},\ \bibinfo {pages} {435} (\bibinfo {year} {1968})}\BibitemShut
  {NoStop}%
\bibitem [{\citenamefont {Rusinov}(1969)}]{rusinov.69}%
  \BibitemOpen
  \bibfield  {author} {\bibinfo {author} {\bibfnamefont {A.~I.}\ \bibnamefont
  {Rusinov}},\ }\href@noop {} {\bibfield  {journal} {\bibinfo  {journal} {Sov.
  JETP Lett.}\ }\textbf {\bibinfo {volume} {9}},\ \bibinfo {pages} {85}
  (\bibinfo {year} {1969})}\BibitemShut {NoStop}%
\bibitem [{\citenamefont {Yazdani}\ \emph {et~al.}(1997)\citenamefont
  {Yazdani}, \citenamefont {Jones}, \citenamefont {Lutz}, \citenamefont
  {Crommie},\ and\ \citenamefont {Eigler}}]{yazdani.jones.97}%
  \BibitemOpen
  \bibfield  {author} {\bibinfo {author} {\bibfnamefont {A.}~\bibnamefont
  {Yazdani}}, \bibinfo {author} {\bibfnamefont {B.~A.}\ \bibnamefont {Jones}},
  \bibinfo {author} {\bibfnamefont {C.~P.}\ \bibnamefont {Lutz}}, \bibinfo
  {author} {\bibfnamefont {M.~F.}\ \bibnamefont {Crommie}}, \ and\ \bibinfo
  {author} {\bibfnamefont {D.~M.}\ \bibnamefont {Eigler}},\ }\bibfield  {title}
  {\enquote {\bibinfo {title} {Probing the local effects of magnetic impurities
  on superconductivity},}\ }\href {\doibase 10.1126/science.275.5307.1767}
  {\bibfield  {journal} {\bibinfo  {journal} {Science}\ }\textbf {\bibinfo
  {volume} {275}},\ \bibinfo {pages} {1767} (\bibinfo {year}
  {1997})}\BibitemShut {NoStop}%
\bibitem [{\citenamefont {Ji}\ \emph {et~al.}(2008)\citenamefont {Ji},
  \citenamefont {Zhang}, \citenamefont {Fu}, \citenamefont {Chen},
  \citenamefont {Ma}, \citenamefont {Li}, \citenamefont {Duan}, \citenamefont
  {Jia},\ and\ \citenamefont {Xue}}]{ji.zhang.08}%
  \BibitemOpen
  \bibfield  {author} {\bibinfo {author} {\bibfnamefont {S.-H.}\ \bibnamefont
  {Ji}}, \bibinfo {author} {\bibfnamefont {T.}~\bibnamefont {Zhang}}, \bibinfo
  {author} {\bibfnamefont {Y.-S.}\ \bibnamefont {Fu}}, \bibinfo {author}
  {\bibfnamefont {X.}~\bibnamefont {Chen}}, \bibinfo {author} {\bibfnamefont
  {X.-C.}\ \bibnamefont {Ma}}, \bibinfo {author} {\bibfnamefont
  {J.}~\bibnamefont {Li}}, \bibinfo {author} {\bibfnamefont {W.-H.}\
  \bibnamefont {Duan}}, \bibinfo {author} {\bibfnamefont {J-F.}\ \bibnamefont
  {Jia}}, \ and\ \bibinfo {author} {\bibfnamefont {Q.-K.}\ \bibnamefont
  {Xue}},\ }\bibfield  {title} {\enquote {\bibinfo {title} {High-resolution
  scanning tunneling spectroscopy of magnetic impurity induced bound states in
  the superconducting gap of {Pb} thin films},}\ }\href {\doibase
  10.1103/PhysRevLett.100.226801} {\bibfield  {journal} {\bibinfo  {journal}
  {Phys. Rev. Lett.}\ }\textbf {\bibinfo {volume} {100}},\ \bibinfo {pages}
  {226801} (\bibinfo {year} {2008})}\BibitemShut {NoStop}%
\bibitem [{\citenamefont {Ruby}\ \emph {et~al.}(2015)\citenamefont {Ruby},
  \citenamefont {Pientka}, \citenamefont {Peng}, \citenamefont {von Oppen},
  \citenamefont {Heinrich},\ and\ \citenamefont {Franke}}]{ruby.pientka.15}%
  \BibitemOpen
  \bibfield  {author} {\bibinfo {author} {\bibfnamefont {M.}~\bibnamefont
  {Ruby}}, \bibinfo {author} {\bibfnamefont {F.}~\bibnamefont {Pientka}},
  \bibinfo {author} {\bibfnamefont {Y.}~\bibnamefont {Peng}}, \bibinfo {author}
  {\bibfnamefont {F.}~\bibnamefont {von Oppen}}, \bibinfo {author}
  {\bibfnamefont {B.~W.}\ \bibnamefont {Heinrich}}, \ and\ \bibinfo {author}
  {\bibfnamefont {K.~J.}\ \bibnamefont {Franke}},\ }\bibfield  {title}
  {\enquote {\bibinfo {title} {Tunneling processes into localized subgap states
  in superconductors},}\ }\href {\doibase 10.1103/PhysRevLett.115.087001}
  {\bibfield  {journal} {\bibinfo  {journal} {Phys. Rev. Lett.}\ }\textbf
  {\bibinfo {volume} {115}},\ \bibinfo {pages} {087001} (\bibinfo {year}
  {2015})}\BibitemShut {NoStop}%
\bibitem [{\citenamefont {Hatter}\ \emph {et~al.}(2015)\citenamefont {Hatter},
  \citenamefont {Heinrich}, \citenamefont {Ruby}, \citenamefont {Pascual},\
  and\ \citenamefont {Franke}}]{hatter.heinrich.15}%
  \BibitemOpen
  \bibfield  {author} {\bibinfo {author} {\bibfnamefont {N.}~\bibnamefont
  {Hatter}}, \bibinfo {author} {\bibfnamefont {B.~W.}\ \bibnamefont
  {Heinrich}}, \bibinfo {author} {\bibfnamefont {M.}~\bibnamefont {Ruby}},
  \bibinfo {author} {\bibfnamefont {J.~I.}\ \bibnamefont {Pascual}}, \ and\
  \bibinfo {author} {\bibfnamefont {K.~J.}\ \bibnamefont {Franke}},\ }\bibfield
   {title} {\enquote {\bibinfo {title} {Magnetic anisotropy in {Shiba} bound
  states across a quantum phase transition},}\ }\href {\doibase
  10.1038/ncomms9988} {\bibfield  {journal} {\bibinfo  {journal} {Nat.
  Commun.}\ }\textbf {\bibinfo {volume} {6}},\ \bibinfo {pages} {8988}
  (\bibinfo {year} {2015})}\BibitemShut {NoStop}%
\bibitem [{\citenamefont {M\'{e}nard}\ \emph {et~al.}(2015)\citenamefont
  {M\'{e}nard}, \citenamefont {Guissart}, \citenamefont {Brun}, \citenamefont
  {Pons}, \citenamefont {Stolyarov}, \citenamefont {Debontridder},
  \citenamefont {Leclerc}, \citenamefont {Janod}, \citenamefont {Cario},
  \citenamefont {Roditchev}, \citenamefont {Simon},\ and\ \citenamefont
  {Cren}}]{menard.guissart.15}%
  \BibitemOpen
  \bibfield  {author} {\bibinfo {author} {\bibfnamefont {G.~C.}\ \bibnamefont
  {M\'{e}nard}}, \bibinfo {author} {\bibfnamefont {S.}~\bibnamefont
  {Guissart}}, \bibinfo {author} {\bibfnamefont {C.}~\bibnamefont {Brun}},
  \bibinfo {author} {\bibfnamefont {S.}~\bibnamefont {Pons}}, \bibinfo {author}
  {\bibfnamefont {V.~S.}\ \bibnamefont {Stolyarov}}, \bibinfo {author}
  {\bibfnamefont {F.}~\bibnamefont {Debontridder}}, \bibinfo {author}
  {\bibfnamefont {M.~V.}\ \bibnamefont {Leclerc}}, \bibinfo {author}
  {\bibfnamefont {E.}~\bibnamefont {Janod}}, \bibinfo {author} {\bibfnamefont
  {L.}~\bibnamefont {Cario}}, \bibinfo {author} {\bibfnamefont
  {D.}~\bibnamefont {Roditchev}}, \bibinfo {author} {\bibfnamefont
  {P.}~\bibnamefont {Simon}}, \ and\ \bibinfo {author} {\bibfnamefont
  {T.}~\bibnamefont {Cren}},\ }\bibfield  {title} {\enquote {\bibinfo {title}
  {Coherent long-range magnetic bound states in a superconductor},}\ }\href
  {\doibase 10.1038/nphys3508} {\bibfield  {journal} {\bibinfo  {journal} {Nat.
  Phys.}\ }\textbf {\bibinfo {volume} {11}},\ \bibinfo {pages} {1013} (\bibinfo
  {year} {2015})}\BibitemShut {NoStop}%
\bibitem [{\citenamefont {Choi}\ \emph {et~al.}(2017)\citenamefont {Choi},
  \citenamefont {Rubio-Verd\'{u}}, \citenamefont {de~Bruijckere}, \citenamefont
  {Ugeda}, \citenamefont {Lorente},\ and\ \citenamefont
  {Pascual}}]{choi.rubioverdu.17}%
  \BibitemOpen
  \bibfield  {author} {\bibinfo {author} {\bibfnamefont {D.-J.}\ \bibnamefont
  {Choi}}, \bibinfo {author} {\bibfnamefont {C.}~\bibnamefont
  {Rubio-Verd\'{u}}}, \bibinfo {author} {\bibfnamefont {J.}~\bibnamefont
  {de~Bruijckere}}, \bibinfo {author} {\bibfnamefont {M.~M.}\ \bibnamefont
  {Ugeda}}, \bibinfo {author} {\bibfnamefont {N.}~\bibnamefont {Lorente}}, \
  and\ \bibinfo {author} {\bibfnamefont {J.~I.}\ \bibnamefont {Pascual}},\
  }\bibfield  {title} {\enquote {\bibinfo {title} {Mapping the orbital
  structure of impurity bound states in a superconductor},}\ }\href {\doibase
  10.1038/ncomms15175} {\bibfield  {journal} {\bibinfo  {journal} {Nat.
  Commun.}\ }\textbf {\bibinfo {volume} {8}},\ \bibinfo {pages} {15175}
  (\bibinfo {year} {2017})}\BibitemShut {NoStop}%
\bibitem [{\citenamefont {Assouline}\ \emph {et~al.}(2017)\citenamefont
  {Assouline}, \citenamefont {Feuillet-Palma}, \citenamefont {Zimmers},
  \citenamefont {Aubin}, \citenamefont {Aprili},\ and\ \citenamefont
  {Harmand}}]{assouline.feuilletpalma.17}%
  \BibitemOpen
  \bibfield  {author} {\bibinfo {author} {\bibfnamefont {A.}~\bibnamefont
  {Assouline}}, \bibinfo {author} {\bibfnamefont {Ch.}\ \bibnamefont
  {Feuillet-Palma}}, \bibinfo {author} {\bibfnamefont {A.}~\bibnamefont
  {Zimmers}}, \bibinfo {author} {\bibfnamefont {Herv\'e}\ \bibnamefont
  {Aubin}}, \bibinfo {author} {\bibfnamefont {M.}~\bibnamefont {Aprili}}, \
  and\ \bibinfo {author} {\bibfnamefont {J.-Ch.}\ \bibnamefont {Harmand}},\
  }\bibfield  {title} {\enquote {\bibinfo {title} {Shiba bound states across
  the mobility edge in doped {InAs} nanowires},}\ }\href {\doibase
  10.1103/PhysRevLett.119.097701} {\bibfield  {journal} {\bibinfo  {journal}
  {Phys. Rev. Lett.}\ }\textbf {\bibinfo {volume} {119}},\ \bibinfo {pages}
  {097701} (\bibinfo {year} {2017})}\BibitemShut {NoStop}%
\bibitem [{\citenamefont {Jellinggaard}\ \emph {et~al.}(2016)\citenamefont
  {Jellinggaard}, \citenamefont {Grove-Rasmussen}, \citenamefont {Madsen},\
  and\ \citenamefont {Nyg\aa{}rd}}]{jellinggaard.groverasmussen.16}%
  \BibitemOpen
  \bibfield  {author} {\bibinfo {author} {\bibfnamefont {A.}~\bibnamefont
  {Jellinggaard}}, \bibinfo {author} {\bibfnamefont {K.}~\bibnamefont
  {Grove-Rasmussen}}, \bibinfo {author} {\bibfnamefont {M.~H.}\ \bibnamefont
  {Madsen}}, \ and\ \bibinfo {author} {\bibfnamefont {J.}~\bibnamefont
  {Nyg\aa{}rd}},\ }\bibfield  {title} {\enquote {\bibinfo {title} {Tuning
  {Yu-Shiba-Rusinov} states in a quantum dot},}\ }\href {\doibase
  10.1103/PhysRevB.94.064520} {\bibfield  {journal} {\bibinfo  {journal} {Phys.
  Rev. B}\ }\textbf {\bibinfo {volume} {94}},\ \bibinfo {pages} {064520}
  (\bibinfo {year} {2016})}\BibitemShut {NoStop}%
\bibitem [{\citenamefont {Salkola}\ \emph {et~al.}(1997)\citenamefont
  {Salkola}, \citenamefont {Balatsky},\ and\ \citenamefont
  {Schrieffer}}]{salkola.balatsky.97}%
  \BibitemOpen
  \bibfield  {author} {\bibinfo {author} {\bibfnamefont {M.~I.}\ \bibnamefont
  {Salkola}}, \bibinfo {author} {\bibfnamefont {A.~V.}\ \bibnamefont
  {Balatsky}}, \ and\ \bibinfo {author} {\bibfnamefont {J.~R.}\ \bibnamefont
  {Schrieffer}},\ }\bibfield  {title} {\enquote {\bibinfo {title} {Spectral
  properties of quasiparticle excitations induced by magnetic moments in
  superconductors},}\ }\href {\doibase 10.1103/PhysRevB.55.12648} {\bibfield
  {journal} {\bibinfo  {journal} {Phys. Rev. B}\ }\textbf {\bibinfo {volume}
  {55}},\ \bibinfo {pages} {12648} (\bibinfo {year} {1997})}\BibitemShut
  {NoStop}%
\bibitem [{\citenamefont {Flatt\'{e}}\ and\ \citenamefont
  {Byers}(1997)}]{flatte.byers.97}%
  \BibitemOpen
  \bibfield  {author} {\bibinfo {author} {\bibfnamefont {M.~E.}\ \bibnamefont
  {Flatt\'{e}}}\ and\ \bibinfo {author} {\bibfnamefont {J.~M.}\ \bibnamefont
  {Byers}},\ }\bibfield  {title} {\enquote {\bibinfo {title} {Local electronic
  structure of a single magnetic impurity in a superconductor},}\ }\href
  {\doibase 10.1103/PhysRevLett.78.3761} {\bibfield  {journal} {\bibinfo
  {journal} {Phys. Rev. Lett.}\ }\textbf {\bibinfo {volume} {78}},\ \bibinfo
  {pages} {3761} (\bibinfo {year} {1997})}\BibitemShut {NoStop}%
\bibitem [{\citenamefont {Fetter}(1965)}]{fetter.65}%
  \BibitemOpen
  \bibfield  {author} {\bibinfo {author} {\bibfnamefont {A.~L.}\ \bibnamefont
  {Fetter}},\ }\bibfield  {title} {\enquote {\bibinfo {title} {Spherical
  impurity in an infinite superconductor},}\ }\href {\doibase
  10.1103/PhysRev.140.A1921} {\bibfield  {journal} {\bibinfo  {journal} {Phys.
  Rev.}\ }\textbf {\bibinfo {volume} {140}},\ \bibinfo {pages} {A1921}
  (\bibinfo {year} {1965})}\BibitemShut {NoStop}%
\bibitem [{\citenamefont {Ugeda}\ \emph {et~al.}(2016)\citenamefont {Ugeda},
  \citenamefont {Bradley}, \citenamefont {Zhang}, \citenamefont {Onishi},
  \citenamefont {Chen}, \citenamefont {Ruan}, \citenamefont
  {Ojeda-Aristizabal}, \citenamefont {Ryu}, \citenamefont {Edmonds},
  \citenamefont {Tsai}, \citenamefont {Riss}, \citenamefont {Mo}, \citenamefont
  {Lee}, \citenamefont {Zettl}, \citenamefont {Hussain}, \citenamefont {Shen},\
  and\ \citenamefont {Crommie}}]{ugeda.bradley.16}%
  \BibitemOpen
  \bibfield  {author} {\bibinfo {author} {\bibfnamefont {M.~M.}\ \bibnamefont
  {Ugeda}}, \bibinfo {author} {\bibfnamefont {A.~J.}\ \bibnamefont {Bradley}},
  \bibinfo {author} {\bibfnamefont {Y.}~\bibnamefont {Zhang}}, \bibinfo
  {author} {\bibfnamefont {S.}~\bibnamefont {Onishi}}, \bibinfo {author}
  {\bibfnamefont {Y.}~\bibnamefont {Chen}}, \bibinfo {author} {\bibfnamefont
  {W.}~\bibnamefont {Ruan}}, \bibinfo {author} {\bibfnamefont {C.}~\bibnamefont
  {Ojeda-Aristizabal}}, \bibinfo {author} {\bibfnamefont {H.}~\bibnamefont
  {Ryu}}, \bibinfo {author} {\bibfnamefont {M.~T.}\ \bibnamefont {Edmonds}},
  \bibinfo {author} {\bibfnamefont {H.-Z.}\ \bibnamefont {Tsai}}, \bibinfo
  {author} {\bibfnamefont {A.}~\bibnamefont {Riss}}, \bibinfo {author}
  {\bibfnamefont {S.-K.}\ \bibnamefont {Mo}}, \bibinfo {author} {\bibfnamefont
  {D.}~\bibnamefont {Lee}}, \bibinfo {author} {\bibfnamefont {A.}~\bibnamefont
  {Zettl}}, \bibinfo {author} {\bibfnamefont {Z.}~\bibnamefont {Hussain}},
  \bibinfo {author} {\bibfnamefont {Z.-X.}\ \bibnamefont {Shen}}, \ and\
  \bibinfo {author} {\bibfnamefont {M.~F.}\ \bibnamefont {Crommie}},\
  }\bibfield  {title} {\enquote {\bibinfo {title} {Characterization of
  collective ground states in single-layer {NbSe$_{2}$}},}\ }\href
  {http://doi.org/10.1038/nphys3527} {\bibfield  {journal} {\bibinfo  {journal}
  {Nat. Phys.}\ }\textbf {\bibinfo {volume} {12}},\ \bibinfo {pages} {92}
  (\bibinfo {year} {2016})}\BibitemShut {NoStop}%
\bibitem [{\citenamefont {Xi}\ \emph {et~al.}(2016)\citenamefont {Xi},
  \citenamefont {Wang}, \citenamefont {Zhao}, \citenamefont {Park},
  \citenamefont {Law}, \citenamefont {Berger}, \citenamefont {Forro},
  \citenamefont {Shan},\ and\ \citenamefont {Mak}}]{xi.wang.16}%
  \BibitemOpen
  \bibfield  {author} {\bibinfo {author} {\bibfnamefont {X.}~\bibnamefont
  {Xi}}, \bibinfo {author} {\bibfnamefont {Z.}~\bibnamefont {Wang}}, \bibinfo
  {author} {\bibfnamefont {W.}~\bibnamefont {Zhao}}, \bibinfo {author}
  {\bibfnamefont {J.-H.}\ \bibnamefont {Park}}, \bibinfo {author}
  {\bibfnamefont {K.~T.}\ \bibnamefont {Law}}, \bibinfo {author} {\bibfnamefont
  {H.}~\bibnamefont {Berger}}, \bibinfo {author} {\bibfnamefont
  {L.}~\bibnamefont {Forro}}, \bibinfo {author} {\bibfnamefont
  {J.}~\bibnamefont {Shan}}, \ and\ \bibinfo {author} {\bibfnamefont {K.~F.}\
  \bibnamefont {Mak}},\ }\bibfield  {title} {\enquote {\bibinfo {title} {Ising
  pairing in superconducting {NbSe$_{2}$} atomic layers},}\ }\href {\doibase
  10.1038/nphys3538} {\bibfield  {journal} {\bibinfo  {journal} {Nat. Phys.}\
  }\textbf {\bibinfo {volume} {12}},\ \bibinfo {pages} {139} (\bibinfo {year}
  {2016})}\BibitemShut {NoStop}%
\bibitem [{\citenamefont {Kezilebieke}\ \emph {et~al.}(2017)\citenamefont
  {Kezilebieke}, \citenamefont {Dvorak}, \citenamefont {Ojanen},\ and\
  \citenamefont {Liljeroth}}]{kezilebieke.dvorak.17}%
  \BibitemOpen
  \bibfield  {author} {\bibinfo {author} {\bibfnamefont {S.}~\bibnamefont
  {Kezilebieke}}, \bibinfo {author} {\bibfnamefont {M.}~\bibnamefont {Dvorak}},
  \bibinfo {author} {\bibfnamefont {T.}~\bibnamefont {Ojanen}}, \ and\ \bibinfo
  {author} {\bibfnamefont {P.}~\bibnamefont {Liljeroth}},\ }\href@noop {}
  {\enquote {\bibinfo {title} {Coupled {Yu-Shiba-Rusinov} states in molecular
  dimers on {NbSe$_2$}},}\ } (\bibinfo {year} {2017}),\ \Eprint
  {http://arxiv.org/abs/arXiv:1701.03288} {arXiv:1701.03288} \BibitemShut
  {NoStop}%
\bibitem [{\citenamefont {Bawden}\ \emph {et~al.}(2016)\citenamefont {Bawden},
  \citenamefont {Cooil}, \citenamefont {Mazzola}, \citenamefont {Riley},
  \citenamefont {Collins-McIntyre}, \citenamefont {Sunko}, \citenamefont
  {Hunvik}, \citenamefont {Leandersson}, \citenamefont {Polley}, \citenamefont
  {Balasubramanian}, \citenamefont {Kim}, \citenamefont {Hoesch}, \citenamefont
  {Wells}, \citenamefont {Balakrishnan}, \citenamefont {Bahramy},\ and\
  \citenamefont {King}}]{bawden.cooil.16}%
  \BibitemOpen
  \bibfield  {author} {\bibinfo {author} {\bibfnamefont {L.}~\bibnamefont
  {Bawden}}, \bibinfo {author} {\bibfnamefont {S.~P.}\ \bibnamefont {Cooil}},
  \bibinfo {author} {\bibfnamefont {F.}~\bibnamefont {Mazzola}}, \bibinfo
  {author} {\bibfnamefont {J.~M.}\ \bibnamefont {Riley}}, \bibinfo {author}
  {\bibfnamefont {L.~J.}\ \bibnamefont {Collins-McIntyre}}, \bibinfo {author}
  {\bibfnamefont {V.}~\bibnamefont {Sunko}}, \bibinfo {author} {\bibfnamefont
  {K.~W.~B.}\ \bibnamefont {Hunvik}}, \bibinfo {author} {\bibfnamefont
  {M.}~\bibnamefont {Leandersson}}, \bibinfo {author} {\bibfnamefont {C.~M.}\
  \bibnamefont {Polley}}, \bibinfo {author} {\bibfnamefont {T.}~\bibnamefont
  {Balasubramanian}}, \bibinfo {author} {\bibfnamefont {T.~K.}\ \bibnamefont
  {Kim}}, \bibinfo {author} {\bibfnamefont {M.}~\bibnamefont {Hoesch}},
  \bibinfo {author} {\bibfnamefont {J.~W.}\ \bibnamefont {Wells}}, \bibinfo
  {author} {\bibfnamefont {G.}~\bibnamefont {Balakrishnan}}, \bibinfo {author}
  {\bibfnamefont {M.~S.}\ \bibnamefont {Bahramy}}, \ and\ \bibinfo {author}
  {\bibfnamefont {P.~D.~C.}\ \bibnamefont {King}},\ }\bibfield  {title}
  {\enquote {\bibinfo {title} {Spin--valley locking in the normal state of a
  transition-metal dichalcogenide superconductor},}\ }\href {\doibase
  10.1038/ncomms11711} {\bibfield  {journal} {\bibinfo  {journal} {Nat.
  Commun.}\ }\textbf {\bibinfo {volume} {7}},\ \bibinfo {pages} {11711}
  (\bibinfo {year} {2016})}\BibitemShut {NoStop}%
\bibitem [{\citenamefont {Meerschaut}\ and\ \citenamefont
  {Deudon}(2001)}]{meerschaut.deudon.01}%
  \BibitemOpen
  \bibfield  {author} {\bibinfo {author} {\bibfnamefont {A.}~\bibnamefont
  {Meerschaut}}\ and\ \bibinfo {author} {\bibfnamefont {C.}~\bibnamefont
  {Deudon}},\ }\bibfield  {title} {\enquote {\bibinfo {title} {Crystal
  structure studies of the {3R-Nb$_{1.09}$S$_{2}$} and the {2H-NbSe$_{2}$}
  compounds: correlation between nonstoichiometry and stacking type (=
  polytypism)},}\ }\href {\doibase 10.1016/S0025-5408(01)00646-8} {\bibfield
  {journal} {\bibinfo  {journal} {Mater. Res. Bull.}\ }\textbf {\bibinfo
  {volume} {36}},\ \bibinfo {pages} {1721} (\bibinfo {year}
  {2001})}\BibitemShut {NoStop}%
\bibitem [{\citenamefont {Riley}\ \emph {et~al.}(2014)\citenamefont {Riley},
  \citenamefont {Mazzola}, \citenamefont {Dendzik}, \citenamefont {Michiardi},
  \citenamefont {Takayama}, \citenamefont {Bawden}, \citenamefont
  {Graner{\o}d}, \citenamefont {Leandersson}, \citenamefont {Balasubramanian},
  \citenamefont {Hoesch}, \citenamefont {Kim}, \citenamefont {Takagi},
  \citenamefont {Meevasana}, \citenamefont {Hofmann}, \citenamefont {Bahramy},
  \citenamefont {Wells},\ and\ \citenamefont {King}}]{riley.mazzola.14}%
  \BibitemOpen
  \bibfield  {author} {\bibinfo {author} {\bibfnamefont {J.~M.}\ \bibnamefont
  {Riley}}, \bibinfo {author} {\bibfnamefont {F.}~\bibnamefont {Mazzola}},
  \bibinfo {author} {\bibfnamefont {M.}~\bibnamefont {Dendzik}}, \bibinfo
  {author} {\bibfnamefont {M.}~\bibnamefont {Michiardi}}, \bibinfo {author}
  {\bibfnamefont {T.}~\bibnamefont {Takayama}}, \bibinfo {author}
  {\bibfnamefont {L.}~\bibnamefont {Bawden}}, \bibinfo {author} {\bibfnamefont
  {C.}~\bibnamefont {Graner{\o}d}}, \bibinfo {author} {\bibfnamefont
  {M.}~\bibnamefont {Leandersson}}, \bibinfo {author} {\bibfnamefont
  {T.}~\bibnamefont {Balasubramanian}}, \bibinfo {author} {\bibfnamefont
  {M.}~\bibnamefont {Hoesch}}, \bibinfo {author} {\bibfnamefont {T.~K.}\
  \bibnamefont {Kim}}, \bibinfo {author} {\bibfnamefont {H.}~\bibnamefont
  {Takagi}}, \bibinfo {author} {\bibfnamefont {W.}~\bibnamefont {Meevasana}},
  \bibinfo {author} {\bibfnamefont {Ph.}\ \bibnamefont {Hofmann}}, \bibinfo
  {author} {\bibfnamefont {M.~S.}\ \bibnamefont {Bahramy}}, \bibinfo {author}
  {\bibfnamefont {J.~W.}\ \bibnamefont {Wells}}, \ and\ \bibinfo {author}
  {\bibfnamefont {P.~D.~C.}\ \bibnamefont {King}},\ }\bibfield  {title}
  {\enquote {\bibinfo {title} {Direct observation of spin-polarised bulk bands
  in an inversion-symmetric semiconductor},}\ }\href {\doibase
  10.1038/nphys3105} {\bibfield  {journal} {\bibinfo  {journal} {Nat. Phys.}\
  }\textbf {\bibinfo {volume} {10}},\ \bibinfo {pages} {835} (\bibinfo {year}
  {2014})}\BibitemShut {NoStop}%
\bibitem [{\citenamefont {Zhang}\ \emph {et~al.}(2014)\citenamefont {Zhang},
  \citenamefont {Liu}, \citenamefont {Luo}, \citenamefont {Freeman},\ and\
  \citenamefont {Zunger}}]{zhang.liu.14}%
  \BibitemOpen
  \bibfield  {author} {\bibinfo {author} {\bibfnamefont {X.}~\bibnamefont
  {Zhang}}, \bibinfo {author} {\bibfnamefont {Q.}~\bibnamefont {Liu}}, \bibinfo
  {author} {\bibfnamefont {J.-W.}\ \bibnamefont {Luo}}, \bibinfo {author}
  {\bibfnamefont {A..~J.}\ \bibnamefont {Freeman}}, \ and\ \bibinfo {author}
  {\bibfnamefont {A.}~\bibnamefont {Zunger}},\ }\bibfield  {title} {\enquote
  {\bibinfo {title} {Hidden spin polarization in inversion-symmetric bulk
  crystals},}\ }\href {\doibase 10.1038/nphys2933} {\bibfield  {journal}
  {\bibinfo  {journal} {Nat. Phys.}\ }\textbf {\bibinfo {volume} {10}},\
  \bibinfo {pages} {387} (\bibinfo {year} {2014})}\BibitemShut {NoStop}%
\bibitem [{\citenamefont {Riley}\ \emph {et~al.}(2015)\citenamefont {Riley},
  \citenamefont {Meevasana}, \citenamefont {Bawden}, \citenamefont {Asakawa},
  \citenamefont {Takayama}, \citenamefont {Eknapakul}, \citenamefont {Kim},
  \citenamefont {Hoesch}, \citenamefont {Mo}, \citenamefont {Takagi},
  \citenamefont {Sasagawa}, \citenamefont {Bahramy},\ and\ \citenamefont
  {King}}]{riley.meevasana.15}%
  \BibitemOpen
  \bibfield  {author} {\bibinfo {author} {\bibfnamefont {J.~M.}\ \bibnamefont
  {Riley}}, \bibinfo {author} {\bibfnamefont {W.}~\bibnamefont {Meevasana}},
  \bibinfo {author} {\bibfnamefont {L.}~\bibnamefont {Bawden}}, \bibinfo
  {author} {\bibfnamefont {M.}~\bibnamefont {Asakawa}}, \bibinfo {author}
  {\bibfnamefont {T.}~\bibnamefont {Takayama}}, \bibinfo {author}
  {\bibfnamefont {T.}~\bibnamefont {Eknapakul}}, \bibinfo {author}
  {\bibfnamefont {T.~K.}\ \bibnamefont {Kim}}, \bibinfo {author} {\bibfnamefont
  {M.}~\bibnamefont {Hoesch}}, \bibinfo {author} {\bibfnamefont {S.~K.}\
  \bibnamefont {Mo}}, \bibinfo {author} {\bibfnamefont {H.}~\bibnamefont
  {Takagi}}, \bibinfo {author} {\bibfnamefont {T.}~\bibnamefont {Sasagawa}},
  \bibinfo {author} {\bibfnamefont {M.~S.}\ \bibnamefont {Bahramy}}, \ and\
  \bibinfo {author} {\bibfnamefont {P.~D.~C.}\ \bibnamefont {King}},\
  }\bibfield  {title} {\enquote {\bibinfo {title} {Negative electronic
  compressibility and tunable spin splitting in {WSe$_{2}$}},}\ }\href
  {\doibase 10.1038/nnano.2015.217} {\bibfield  {journal} {\bibinfo  {journal}
  {Nat. Nanotechnol.}\ }\textbf {\bibinfo {volume} {10}},\ \bibinfo {pages}
  {1043} (\bibinfo {year} {2015})}\BibitemShut {NoStop}%
\bibitem [{\citenamefont {Wilson}\ \emph {et~al.}(1974)\citenamefont {Wilson},
  \citenamefont {Di~Salvo},\ and\ \citenamefont {Mahajan}}]{wilson.disalvo.74}%
  \BibitemOpen
  \bibfield  {author} {\bibinfo {author} {\bibfnamefont {J.~A.}\ \bibnamefont
  {Wilson}}, \bibinfo {author} {\bibfnamefont {F.~J.}\ \bibnamefont
  {Di~Salvo}}, \ and\ \bibinfo {author} {\bibfnamefont {S.}~\bibnamefont
  {Mahajan}},\ }\bibfield  {title} {\enquote {\bibinfo {title} {Charge-density
  waves in metallic, layered, transition-metal dichalcogenides},}\ }\href
  {\doibase 10.1103/PhysRevLett.32.882} {\bibfield  {journal} {\bibinfo
  {journal} {Phys. Rev. Lett.}\ }\textbf {\bibinfo {volume} {32}},\ \bibinfo
  {pages} {882} (\bibinfo {year} {1974})}\BibitemShut {NoStop}%
\bibitem [{\citenamefont {Zhu}\ \emph {et~al.}(2015)\citenamefont {Zhu},
  \citenamefont {Cao}, \citenamefont {Zhang}, \citenamefont {Plummer},\ and\
  \citenamefont {Guo}}]{zhu.cao.15}%
  \BibitemOpen
  \bibfield  {author} {\bibinfo {author} {\bibfnamefont {X.}~\bibnamefont
  {Zhu}}, \bibinfo {author} {\bibfnamefont {Y.}~\bibnamefont {Cao}}, \bibinfo
  {author} {\bibfnamefont {J.}~\bibnamefont {Zhang}}, \bibinfo {author}
  {\bibfnamefont {E.~W.}\ \bibnamefont {Plummer}}, \ and\ \bibinfo {author}
  {\bibfnamefont {J.}~\bibnamefont {Guo}},\ }\bibfield  {title} {\enquote
  {\bibinfo {title} {Classification of charge density waves based on their
  nature},}\ }\href {\doibase 10.1073/pnas.1424791112} {\bibfield  {journal}
  {\bibinfo  {journal} {Proc. Natl. Acad. Sci. U.S.A.}\ }\textbf {\bibinfo
  {volume} {112}},\ \bibinfo {pages} {2367} (\bibinfo {year}
  {2015})}\BibitemShut {NoStop}%
\bibitem [{\citenamefont {Yokoya}\ \emph {et~al.}(2001)\citenamefont {Yokoya},
  \citenamefont {Kiss}, \citenamefont {Chainani}, \citenamefont {Shin},
  \citenamefont {Nohara},\ and\ \citenamefont {Takagi}}]{yokoya.kiss.01}%
  \BibitemOpen
  \bibfield  {author} {\bibinfo {author} {\bibfnamefont {T.}~\bibnamefont
  {Yokoya}}, \bibinfo {author} {\bibfnamefont {T.}~\bibnamefont {Kiss}},
  \bibinfo {author} {\bibfnamefont {A.}~\bibnamefont {Chainani}}, \bibinfo
  {author} {\bibfnamefont {S.}~\bibnamefont {Shin}}, \bibinfo {author}
  {\bibfnamefont {M.}~\bibnamefont {Nohara}}, \ and\ \bibinfo {author}
  {\bibfnamefont {H.}~\bibnamefont {Takagi}},\ }\bibfield  {title} {\enquote
  {\bibinfo {title} {Fermi surface sheet-dependent superconductivity in
  {2H-NbSe$_{2}$}},}\ }\href {\doibase 10.1126/science.1065068} {\bibfield
  {journal} {\bibinfo  {journal} {Science}\ }\textbf {\bibinfo {volume}
  {294}},\ \bibinfo {pages} {2518} (\bibinfo {year} {2001})}\BibitemShut
  {NoStop}%
\bibitem [{\citenamefont {Borisenko}\ \emph {et~al.}(2009)\citenamefont
  {Borisenko}, \citenamefont {Kordyuk}, \citenamefont {Zabolotnyy},
  \citenamefont {Inosov}, \citenamefont {Evtushinsky}, \citenamefont
  {B\"{u}chner}, \citenamefont {Yaresko}, \citenamefont {Varykhalov},
  \citenamefont {Follath}, \citenamefont {Eberhardt}, \citenamefont {Patthey},\
  and\ \citenamefont {Berger}}]{borisenko.kordyuk.09}%
  \BibitemOpen
  \bibfield  {author} {\bibinfo {author} {\bibfnamefont {S.~V.}\ \bibnamefont
  {Borisenko}}, \bibinfo {author} {\bibfnamefont {A.~A.}\ \bibnamefont
  {Kordyuk}}, \bibinfo {author} {\bibfnamefont {V.~B.}\ \bibnamefont
  {Zabolotnyy}}, \bibinfo {author} {\bibfnamefont {D.~S.}\ \bibnamefont
  {Inosov}}, \bibinfo {author} {\bibfnamefont {D.}~\bibnamefont {Evtushinsky}},
  \bibinfo {author} {\bibfnamefont {B.}~\bibnamefont {B\"{u}chner}}, \bibinfo
  {author} {\bibfnamefont {A.~N.}\ \bibnamefont {Yaresko}}, \bibinfo {author}
  {\bibfnamefont {A.}~\bibnamefont {Varykhalov}}, \bibinfo {author}
  {\bibfnamefont {R.}~\bibnamefont {Follath}}, \bibinfo {author} {\bibfnamefont
  {W.}~\bibnamefont {Eberhardt}}, \bibinfo {author} {\bibfnamefont
  {L.}~\bibnamefont {Patthey}}, \ and\ \bibinfo {author} {\bibfnamefont
  {H.}~\bibnamefont {Berger}},\ }\bibfield  {title} {\enquote {\bibinfo {title}
  {Two energy gaps and {Fermi}-surface ``arcs'' in {NbSe$_{2}$}},}\ }\href
  {\doibase 10.1103/PhysRevLett.102.166402} {\bibfield  {journal} {\bibinfo
  {journal} {Phys. Rev. Lett.}\ }\textbf {\bibinfo {volume} {102}},\ \bibinfo
  {pages} {166402} (\bibinfo {year} {2009})}\BibitemShut {NoStop}%
\bibitem [{\citenamefont {Rahn}\ \emph {et~al.}(2012)\citenamefont {Rahn},
  \citenamefont {Hellmann}, \citenamefont {Kall\"{a}ne}, \citenamefont {Sohrt},
  \citenamefont {Kim}, \citenamefont {Kipp},\ and\ \citenamefont
  {Rossnagel}}]{rahm.hellmann.12}%
  \BibitemOpen
  \bibfield  {author} {\bibinfo {author} {\bibfnamefont {D.~J.}\ \bibnamefont
  {Rahn}}, \bibinfo {author} {\bibfnamefont {S.}~\bibnamefont {Hellmann}},
  \bibinfo {author} {\bibfnamefont {M.}~\bibnamefont {Kall\"{a}ne}}, \bibinfo
  {author} {\bibfnamefont {C.}~\bibnamefont {Sohrt}}, \bibinfo {author}
  {\bibfnamefont {T.~K.}\ \bibnamefont {Kim}}, \bibinfo {author} {\bibfnamefont
  {L.}~\bibnamefont {Kipp}}, \ and\ \bibinfo {author} {\bibfnamefont
  {K.}~\bibnamefont {Rossnagel}},\ }\bibfield  {title} {\enquote {\bibinfo
  {title} {Gaps and kinks in the electronic structure of the superconductor
  {2H-NbSe$_{2}$} from angle-resolved photoemission at 1 {K}},}\ }\href
  {\doibase 10.1103/PhysRevB.85.224532} {\bibfield  {journal} {\bibinfo
  {journal} {Phys. Rev. B}\ }\textbf {\bibinfo {volume} {85}},\ \bibinfo
  {pages} {224532} (\bibinfo {year} {2012})}\BibitemShut {NoStop}%
\bibitem [{\citenamefont {Rossnagel}\ \emph {et~al.}(2001)\citenamefont
  {Rossnagel}, \citenamefont {Seifarth}, \citenamefont {Kipp}, \citenamefont
  {Skibowski}, \citenamefont {Vo\ss{}}, \citenamefont {Kr\"{u}ger},
  \citenamefont {Mazur},\ and\ \citenamefont
  {Pollmann}}]{rossnagel.seifarth.01}%
  \BibitemOpen
  \bibfield  {author} {\bibinfo {author} {\bibfnamefont {K.}~\bibnamefont
  {Rossnagel}}, \bibinfo {author} {\bibfnamefont {O.}~\bibnamefont {Seifarth}},
  \bibinfo {author} {\bibfnamefont {L.}~\bibnamefont {Kipp}}, \bibinfo {author}
  {\bibfnamefont {M.}~\bibnamefont {Skibowski}}, \bibinfo {author}
  {\bibfnamefont {D.}~\bibnamefont {Vo\ss{}}}, \bibinfo {author} {\bibfnamefont
  {P.}~\bibnamefont {Kr\"{u}ger}}, \bibinfo {author} {\bibfnamefont
  {A.}~\bibnamefont {Mazur}}, \ and\ \bibinfo {author} {\bibfnamefont
  {J.}~\bibnamefont {Pollmann}},\ }\bibfield  {title} {\enquote {\bibinfo
  {title} {Fermi surface of {2H-NbSe$_{2}$} and its implications on the
  charge-density-wave mechanism},}\ }\href {\doibase
  10.1103/PhysRevB.64.235119} {\bibfield  {journal} {\bibinfo  {journal} {Phys.
  Rev. B}\ }\textbf {\bibinfo {volume} {64}},\ \bibinfo {pages} {235119}
  (\bibinfo {year} {2001})}\BibitemShut {NoStop}%
\bibitem [{\citenamefont {Johannes}\ \emph {et~al.}(2006)\citenamefont
  {Johannes}, \citenamefont {Mazin},\ and\ \citenamefont
  {Howells}}]{johannes.mazin.06}%
  \BibitemOpen
  \bibfield  {author} {\bibinfo {author} {\bibfnamefont {M.~D.}\ \bibnamefont
  {Johannes}}, \bibinfo {author} {\bibfnamefont {I.~I.}\ \bibnamefont {Mazin}},
  \ and\ \bibinfo {author} {\bibfnamefont {C.~A.}\ \bibnamefont {Howells}},\
  }\bibfield  {title} {\enquote {\bibinfo {title} {Fermi-surface nesting and
  the origin of the charge-density wave in {NbSe$_{2}$}},}\ }\href {\doibase
  10.1103/PhysRevB.73.205102} {\bibfield  {journal} {\bibinfo  {journal} {Phys.
  Rev. B}\ }\textbf {\bibinfo {volume} {73}},\ \bibinfo {pages} {205102}
  (\bibinfo {year} {2006})}\BibitemShut {NoStop}%
\bibitem [{\citenamefont {Flicker}\ and\ \citenamefont {van
  Wezel}(2015)}]{flicker.vanwezel.15}%
  \BibitemOpen
  \bibfield  {author} {\bibinfo {author} {\bibfnamefont {F.}~\bibnamefont
  {Flicker}}\ and\ \bibinfo {author} {\bibfnamefont {J.}~\bibnamefont {van
  Wezel}},\ }\bibfield  {title} {\enquote {\bibinfo {title} {Charge order from
  orbital-dependent coupling evidenced by {NbSe$_{2}$}},}\ }\href {\doibase
  10.1038/ncomms8034} {\bibfield  {journal} {\bibinfo  {journal} {Nat.
  Commun.}\ }\textbf {\bibinfo {volume} {6}},\ \bibinfo {pages} {7034}
  (\bibinfo {year} {2015})}\BibitemShut {NoStop}%
\bibitem [{\citenamefont {Silva-Guill\'{e}n}\ \emph {et~al.}(2016)\citenamefont
  {Silva-Guill\'{e}n}, \citenamefont {Ordej\'{o}n}, \citenamefont {Guinea},\
  and\ \citenamefont {Canadell}}]{silvaguillen.ordejon.16}%
  \BibitemOpen
  \bibfield  {author} {\bibinfo {author} {\bibfnamefont {J.~\'{A}.}\
  \bibnamefont {Silva-Guill\'{e}n}}, \bibinfo {author} {\bibfnamefont
  {P.}~\bibnamefont {Ordej\'{o}n}}, \bibinfo {author} {\bibfnamefont
  {F.}~\bibnamefont {Guinea}}, \ and\ \bibinfo {author} {\bibfnamefont
  {E.}~\bibnamefont {Canadell}},\ }\bibfield  {title} {\enquote {\bibinfo
  {title} {Electronic structure of {2H-NbSe$_{2}$} single-layers in the {CDW}
  state},}\ }\href {\doibase 10.1088/2053-1583/3/3/035028} {\bibfield
  {journal} {\bibinfo  {journal} {2D Mater.}\ }\textbf {\bibinfo {volume}
  {3}},\ \bibinfo {pages} {035028} (\bibinfo {year} {2016})}\BibitemShut
  {NoStop}%
\bibitem [{\citenamefont {Momma}\ and\ \citenamefont
  {Izumi}(2011)}]{momma.izumi.11}%
  \BibitemOpen
  \bibfield  {author} {\bibinfo {author} {\bibfnamefont {K.}~\bibnamefont
  {Momma}}\ and\ \bibinfo {author} {\bibfnamefont {F.}~\bibnamefont {Izumi}},\
  }\bibfield  {title} {\enquote {\bibinfo {title} {{{\it VESTA3} for
  three-dimensional visualization of crystal, volumetric and morphology
  data}},}\ }\href {\doibase 10.1107/S0021889811038970} {\bibfield  {journal}
  {\bibinfo  {journal} {J. Appl. Crystallogr.}\ }\textbf {\bibinfo {volume}
  {44}},\ \bibinfo {pages} {1272} (\bibinfo {year} {2011})}\BibitemShut
  {NoStop}%
\bibitem [{\citenamefont {Zeng}\ \emph {et~al.}(2012)\citenamefont {Zeng},
  \citenamefont {Dai}, \citenamefont {Yao}, \citenamefont {Xiao},\ and\
  \citenamefont {Cui}}]{zeng.dai.12}%
  \BibitemOpen
  \bibfield  {author} {\bibinfo {author} {\bibfnamefont {H.}~\bibnamefont
  {Zeng}}, \bibinfo {author} {\bibfnamefont {J.}~\bibnamefont {Dai}}, \bibinfo
  {author} {\bibfnamefont {W.}~\bibnamefont {Yao}}, \bibinfo {author}
  {\bibfnamefont {D.}~\bibnamefont {Xiao}}, \ and\ \bibinfo {author}
  {\bibfnamefont {X.}~\bibnamefont {Cui}},\ }\bibfield  {title} {\enquote
  {\bibinfo {title} {Valley polarization in {MoS$_{2}$} monolayers by optical
  pumping},}\ }\href {\doibase 10.1038/nnano.2012.95} {\bibfield  {journal}
  {\bibinfo  {journal} {Nat. Nano.}\ }\textbf {\bibinfo {volume} {7}},\
  \bibinfo {pages} {490} (\bibinfo {year} {2012})}\BibitemShut {NoStop}%
\bibitem [{\citenamefont {Mak}\ \emph {et~al.}(2012)\citenamefont {Mak},
  \citenamefont {He}, \citenamefont {Shan},\ and\ \citenamefont
  {Heinz}}]{mak.he.12}%
  \BibitemOpen
  \bibfield  {author} {\bibinfo {author} {\bibfnamefont {K.~F.}\ \bibnamefont
  {Mak}}, \bibinfo {author} {\bibfnamefont {K.}~\bibnamefont {He}}, \bibinfo
  {author} {\bibfnamefont {J.}~\bibnamefont {Shan}}, \ and\ \bibinfo {author}
  {\bibfnamefont {T.~F.}\ \bibnamefont {Heinz}},\ }\bibfield  {title} {\enquote
  {\bibinfo {title} {Control of valley polarization in monolayer {MoS$_{2}$} by
  optical helicity},}\ }\href {\doibase 10.1038/nnano.2012.96} {\bibfield
  {journal} {\bibinfo  {journal} {Nat. Nano.}\ }\textbf {\bibinfo {volume}
  {7}},\ \bibinfo {pages} {494} (\bibinfo {year} {2012})}\BibitemShut {NoStop}%
\bibitem [{\citenamefont {Suzuki}\ \emph {et~al.}(2014)\citenamefont {Suzuki},
  \citenamefont {Sakano}, \citenamefont {Zhang}, \citenamefont {Akashi},
  \citenamefont {Morikawa}, \citenamefont {Harasawa}, \citenamefont {Yaji},
  \citenamefont {Kuroda}, \citenamefont {Miyamoto}, \citenamefont {Okuda},
  \citenamefont {Ishizaka}, \citenamefont {Arita},\ and\ \citenamefont
  {Iwasa}}]{suzuki.sakano.14}%
  \BibitemOpen
  \bibfield  {author} {\bibinfo {author} {\bibfnamefont {R.}~\bibnamefont
  {Suzuki}}, \bibinfo {author} {\bibfnamefont {M.}~\bibnamefont {Sakano}},
  \bibinfo {author} {\bibfnamefont {Y.~J.}\ \bibnamefont {Zhang}}, \bibinfo
  {author} {\bibfnamefont {R.}~\bibnamefont {Akashi}}, \bibinfo {author}
  {\bibfnamefont {D.}~\bibnamefont {Morikawa}}, \bibinfo {author}
  {\bibfnamefont {A.}~\bibnamefont {Harasawa}}, \bibinfo {author}
  {\bibfnamefont {K.}~\bibnamefont {Yaji}}, \bibinfo {author} {\bibfnamefont
  {K.}~\bibnamefont {Kuroda}}, \bibinfo {author} {\bibfnamefont
  {K.}~\bibnamefont {Miyamoto}}, \bibinfo {author} {\bibfnamefont
  {T.}~\bibnamefont {Okuda}}, \bibinfo {author} {\bibfnamefont
  {K.}~\bibnamefont {Ishizaka}}, \bibinfo {author} {\bibfnamefont
  {R.}~\bibnamefont {Arita}}, \ and\ \bibinfo {author} {\bibfnamefont
  {Y.}~\bibnamefont {Iwasa}},\ }\bibfield  {title} {\enquote {\bibinfo {title}
  {Valley-dependent spin polarization in bulk {MoS$_{2}$} with broken inversion
  symmetry},}\ }\href {\doibase 10.1038/nnano.2014.148} {\bibfield  {journal}
  {\bibinfo  {journal} {Nat. Nano.}\ }\textbf {\bibinfo {volume} {9}},\
  \bibinfo {pages} {611} (\bibinfo {year} {2014})}\BibitemShut {NoStop}%
\bibitem [{\citenamefont {Zhu}\ \emph {et~al.}(2014)\citenamefont {Zhu},
  \citenamefont {Zeng}, \citenamefont {Dai}, \citenamefont {Gong},\ and\
  \citenamefont {Cui}}]{zhu.zeng.14}%
  \BibitemOpen
  \bibfield  {author} {\bibinfo {author} {\bibfnamefont {B.}~\bibnamefont
  {Zhu}}, \bibinfo {author} {\bibfnamefont {H.}~\bibnamefont {Zeng}}, \bibinfo
  {author} {\bibfnamefont {J.}~\bibnamefont {Dai}}, \bibinfo {author}
  {\bibfnamefont {Z.}~\bibnamefont {Gong}}, \ and\ \bibinfo {author}
  {\bibfnamefont {X.}~\bibnamefont {Cui}},\ }\bibfield  {title} {\enquote
  {\bibinfo {title} {Anomalously robust valley polarization and valley
  coherence in bilayer ws2},}\ }\href {\doibase 10.1073/pnas.1406960111}
  {\bibfield  {journal} {\bibinfo  {journal} {Proc. Natl. Acad. Sci. U.S.A.}\
  }\textbf {\bibinfo {volume} {111}},\ \bibinfo {pages} {11606} (\bibinfo
  {year} {2014})}\BibitemShut {NoStop}%
\bibitem [{\citenamefont {Glass}\ \emph {et~al.}(2015)\citenamefont {Glass},
  \citenamefont {Li}, \citenamefont {Adler}, \citenamefont {Aulbach},
  \citenamefont {Fleszar}, \citenamefont {Thomale}, \citenamefont {Hanke},
  \citenamefont {Claessen},\ and\ \citenamefont {Sch\"{a}fer}}]{glass.li.15}%
  \BibitemOpen
  \bibfield  {author} {\bibinfo {author} {\bibfnamefont {S.}~\bibnamefont
  {Glass}}, \bibinfo {author} {\bibfnamefont {G.}~\bibnamefont {Li}}, \bibinfo
  {author} {\bibfnamefont {F.}~\bibnamefont {Adler}}, \bibinfo {author}
  {\bibfnamefont {J.}~\bibnamefont {Aulbach}}, \bibinfo {author} {\bibfnamefont
  {A.}~\bibnamefont {Fleszar}}, \bibinfo {author} {\bibfnamefont
  {R.}~\bibnamefont {Thomale}}, \bibinfo {author} {\bibfnamefont
  {W.}~\bibnamefont {Hanke}}, \bibinfo {author} {\bibfnamefont
  {R.}~\bibnamefont {Claessen}}, \ and\ \bibinfo {author} {\bibfnamefont
  {J.}~\bibnamefont {Sch\"{a}fer}},\ }\bibfield  {title} {\enquote {\bibinfo
  {title} {Triangular spin-orbit-coupled lattice with strong coulomb
  correlations: {Sn} atoms on a {SiC(0001)} substrate},}\ }\href {\doibase
  10.1103/PhysRevLett.114.247602} {\bibfield  {journal} {\bibinfo  {journal}
  {Phys. Rev. Lett.}\ }\textbf {\bibinfo {volume} {114}},\ \bibinfo {pages}
  {247602} (\bibinfo {year} {2015})}\BibitemShut {NoStop}%
\bibitem [{\citenamefont {Sharma}\ and\ \citenamefont
  {Tewari}(2016)}]{sharma.tewari.16}%
  \BibitemOpen
  \bibfield  {author} {\bibinfo {author} {\bibfnamefont {G.}~\bibnamefont
  {Sharma}}\ and\ \bibinfo {author} {\bibfnamefont {S.}~\bibnamefont
  {Tewari}},\ }\bibfield  {title} {\enquote {\bibinfo {title}
  {{Yu-Shiba-Rusinov} states and topological superconductivity in {Ising}
  paired superconductors},}\ }\href {\doibase 10.1103/PhysRevB.94.094515}
  {\bibfield  {journal} {\bibinfo  {journal} {Phys. Rev. B}\ }\textbf {\bibinfo
  {volume} {94}},\ \bibinfo {pages} {094515} (\bibinfo {year}
  {2016})}\BibitemShut {NoStop}%
\bibitem [{\citenamefont {Smith}\ \emph {et~al.}(2016)\citenamefont {Smith},
  \citenamefont {Tanaka},\ and\ \citenamefont {Nagai}}]{smith.tanaka.16}%
  \BibitemOpen
  \bibfield  {author} {\bibinfo {author} {\bibfnamefont {E.~D.~B.}\
  \bibnamefont {Smith}}, \bibinfo {author} {\bibfnamefont {K.}~\bibnamefont
  {Tanaka}}, \ and\ \bibinfo {author} {\bibfnamefont {Y.}~\bibnamefont
  {Nagai}},\ }\bibfield  {title} {\enquote {\bibinfo {title} {Manifestation of
  chirality in the vortex lattice in a two-dimensional topological
  superconductor},}\ }\href {\doibase 10.1103/PhysRevB.94.064515} {\bibfield
  {journal} {\bibinfo  {journal} {Phys. Rev. B}\ }\textbf {\bibinfo {volume}
  {94}},\ \bibinfo {pages} {064515} (\bibinfo {year} {2016})}\BibitemShut
  {NoStop}%
\bibitem [{\citenamefont {Goertzen}\ \emph {et~al.}(2017)\citenamefont
  {Goertzen}, \citenamefont {Tanaka},\ and\ \citenamefont
  {Nagai}}]{goertzen.tanaka.17}%
  \BibitemOpen
  \bibfield  {author} {\bibinfo {author} {\bibfnamefont {S.~L.}\ \bibnamefont
  {Goertzen}}, \bibinfo {author} {\bibfnamefont {K.}~\bibnamefont {Tanaka}}, \
  and\ \bibinfo {author} {\bibfnamefont {Y.}~\bibnamefont {Nagai}},\ }\bibfield
   {title} {\enquote {\bibinfo {title} {Self-consistent study of {Abelian} and
  {non-Abelian} order in a two-dimensional topological superconductor},}\
  }\href {\doibase 10.1103/PhysRevB.95.064509} {\bibfield  {journal} {\bibinfo
  {journal} {Phys. Rev. B}\ }\textbf {\bibinfo {volume} {95}},\ \bibinfo
  {pages} {064509} (\bibinfo {year} {2017})}\BibitemShut {NoStop}%
\bibitem [{\citenamefont {Xu}\ \emph {et~al.}(2014)\citenamefont {Xu},
  \citenamefont {Qu}, \citenamefont {Gong},\ and\ \citenamefont
  {Zhang}}]{xu.qu.14}%
  \BibitemOpen
  \bibfield  {author} {\bibinfo {author} {\bibfnamefont {Y.}~\bibnamefont
  {Xu}}, \bibinfo {author} {\bibfnamefont {Ch.}\ \bibnamefont {Qu}}, \bibinfo
  {author} {\bibfnamefont {M.}~\bibnamefont {Gong}}, \ and\ \bibinfo {author}
  {\bibfnamefont {Ch.}\ \bibnamefont {Zhang}},\ }\bibfield  {title} {\enquote
  {\bibinfo {title} {Competing superfluid orders in spin-orbit-coupled
  fermionic cold-atom optical lattices},}\ }\href {\doibase
  10.1103/PhysRevA.89.013607} {\bibfield  {journal} {\bibinfo  {journal} {Phys.
  Rev. A}\ }\textbf {\bibinfo {volume} {89}},\ \bibinfo {pages} {013607}
  (\bibinfo {year} {2014})}\BibitemShut {NoStop}%
\bibitem [{\citenamefont {Ptok}\ and\ \citenamefont
  {Kapcia}(2015)}]{ptok.kapcia.15}%
  \BibitemOpen
  \bibfield  {author} {\bibinfo {author} {\bibfnamefont {A.}~\bibnamefont
  {Ptok}}\ and\ \bibinfo {author} {\bibfnamefont {K.~J.}\ \bibnamefont
  {Kapcia}},\ }\bibfield  {title} {\enquote {\bibinfo {title} {Probe-type of
  superconductivity by impurity in materials with short coherence length: the
  s- wave and $\eta$-wave phases study},}\ }\href {\doibase
  10.1088/0953-2048/28/4/045022} {\bibfield  {journal} {\bibinfo  {journal}
  {Supercond. Sci. Technol.}\ }\textbf {\bibinfo {volume} {28}},\ \bibinfo
  {pages} {045022} (\bibinfo {year} {2015})}\BibitemShut {NoStop}%
\bibitem [{\citenamefont {De~Gennes}(1999)}]{degennes.99}%
  \BibitemOpen
  \bibfield  {author} {\bibinfo {author} {\bibfnamefont {P.~G.}\ \bibnamefont
  {De~Gennes}},\ }\href@noop {} {\emph {\bibinfo {title} {Superconductivity Of
  Metals And Alloys}}},\ Advanced Books Classics Series\ (\bibinfo  {publisher}
  {Westview Press},\ \bibinfo {year} {1999})\BibitemShut {NoStop}%
\bibitem [{\citenamefont {Matsui}\ \emph {et~al.}(2003)\citenamefont {Matsui},
  \citenamefont {Sato}, \citenamefont {Takahashi}, \citenamefont {Wang},
  \citenamefont {Yang}, \citenamefont {Ding}, \citenamefont {Fujii},
  \citenamefont {Watanabe},\ and\ \citenamefont {Matsuda}}]{matsui.sato.03}%
  \BibitemOpen
  \bibfield  {author} {\bibinfo {author} {\bibfnamefont {H.}~\bibnamefont
  {Matsui}}, \bibinfo {author} {\bibfnamefont {T.}~\bibnamefont {Sato}},
  \bibinfo {author} {\bibfnamefont {T.}~\bibnamefont {Takahashi}}, \bibinfo
  {author} {\bibfnamefont {S.-C.}\ \bibnamefont {Wang}}, \bibinfo {author}
  {\bibfnamefont {H.-B.}\ \bibnamefont {Yang}}, \bibinfo {author}
  {\bibfnamefont {H.}~\bibnamefont {Ding}}, \bibinfo {author} {\bibfnamefont
  {T.}~\bibnamefont {Fujii}}, \bibinfo {author} {\bibfnamefont
  {T.}~\bibnamefont {Watanabe}}, \ and\ \bibinfo {author} {\bibfnamefont
  {A.}~\bibnamefont {Matsuda}},\ }\bibfield  {title} {\enquote {\bibinfo
  {title} {{BCS}-like bogoliubov quasiparticles in high-{T$_{c}$}
  superconductors observed by angle-resolved photoemission spectroscopy},}\
  }\href {\doibase 10.1103/PhysRevLett.90.217002} {\bibfield  {journal}
  {\bibinfo  {journal} {Phys. Rev. Lett.}\ }\textbf {\bibinfo {volume} {90}},\
  \bibinfo {pages} {217002} (\bibinfo {year} {2003})}\BibitemShut {NoStop}%
\bibitem [{sm.(pdf)}]{sm.pdf}%
  \BibitemOpen
  \href@noop {} {} (\bibinfo {year} {pdf}),\ \bibinfo {note} {see Supplemental
  Material at [URL will be inserted by publisher] for compare with results for
  out-of-plane spin orbit coupling.}\BibitemShut {Stop}%
\bibitem [{\citenamefont {Sakurai}(1970)}]{sakurai.70}%
  \BibitemOpen
  \bibfield  {author} {\bibinfo {author} {\bibfnamefont {A.}~\bibnamefont
  {Sakurai}},\ }\bibfield  {title} {\enquote {\bibinfo {title} {Comments on
  superconductors with magnetic impurities},}\ }\href {\doibase
  10.1143/PTP.44.1472} {\bibfield  {journal} {\bibinfo  {journal} {Prog. Theor.
  Phys.}\ }\textbf {\bibinfo {volume} {44}},\ \bibinfo {pages} {1472} (\bibinfo
  {year} {1970})}\BibitemShut {NoStop}%
\bibitem [{\citenamefont {van Gerven~Oei}\ \emph {et~al.}(2017)\citenamefont
  {van Gerven~Oei}, \citenamefont {Tanaskovi\'{c}},\ and\ \citenamefont
  {\v{Z}itko}}]{vangervenoei.tanaskovic.17}%
  \BibitemOpen
  \bibfield  {author} {\bibinfo {author} {\bibfnamefont {W.~V.}\ \bibnamefont
  {van Gerven~Oei}}, \bibinfo {author} {\bibfnamefont {D.}~\bibnamefont
  {Tanaskovi\'{c}}}, \ and\ \bibinfo {author} {\bibfnamefont {R.}~\bibnamefont
  {\v{Z}itko}},\ }\bibfield  {title} {\enquote {\bibinfo {title} {Magnetic
  impurities in spin-split superconductors},}\ }\href {\doibase
  10.1103/PhysRevB.95.085115} {\bibfield  {journal} {\bibinfo  {journal} {Phys.
  Rev. B}\ }\textbf {\bibinfo {volume} {95}},\ \bibinfo {pages} {085115}
  (\bibinfo {year} {2017})}\BibitemShut {NoStop}%
\bibitem [{\citenamefont {Franke}\ \emph {et~al.}(2011)\citenamefont {Franke},
  \citenamefont {Schulze},\ and\ \citenamefont {Pascual}}]{franke.schulze.11}%
  \BibitemOpen
  \bibfield  {author} {\bibinfo {author} {\bibfnamefont {K.~J.}\ \bibnamefont
  {Franke}}, \bibinfo {author} {\bibfnamefont {G.}~\bibnamefont {Schulze}}, \
  and\ \bibinfo {author} {\bibfnamefont {J.~I.}\ \bibnamefont {Pascual}},\
  }\bibfield  {title} {\enquote {\bibinfo {title} {Competition of
  superconducting phenomena and {Kondo} screening at the nanoscale},}\ }\href
  {\doibase 10.1126/science.1202204} {\bibfield  {journal} {\bibinfo  {journal}
  {Science}\ }\textbf {\bibinfo {volume} {332}},\ \bibinfo {pages} {940}
  (\bibinfo {year} {2011})}\BibitemShut {NoStop}%
\bibitem [{\citenamefont {Kaladzhyan}\ \emph {et~al.}(2016)\citenamefont
  {Kaladzhyan}, \citenamefont {Bena},\ and\ \citenamefont
  {Simon}}]{kaladzhyan.bena.16}%
  \BibitemOpen
  \bibfield  {author} {\bibinfo {author} {\bibfnamefont {V.}~\bibnamefont
  {Kaladzhyan}}, \bibinfo {author} {\bibfnamefont {C.}~\bibnamefont {Bena}}, \
  and\ \bibinfo {author} {\bibfnamefont {P.}~\bibnamefont {Simon}},\ }\bibfield
   {title} {\enquote {\bibinfo {title} {Characterizing $p$-wave
  superconductivity using the spin structure of {Shiba} states},}\ }\href
  {\doibase 10.1103/PhysRevB.93.214514} {\bibfield  {journal} {\bibinfo
  {journal} {Phys. Rev. B}\ }\textbf {\bibinfo {volume} {93}},\ \bibinfo
  {pages} {214514} (\bibinfo {year} {2016})}\BibitemShut {NoStop}%
\bibitem [{\citenamefont {Mashkoori}\ \emph {et~al.}(2017)\citenamefont
  {Mashkoori}, \citenamefont {Bj\"{o}rnson},\ and\ \citenamefont
  {Black-Schaffer}}]{mashkoori.bjornson.17}%
  \BibitemOpen
  \bibfield  {author} {\bibinfo {author} {\bibfnamefont {M.}~\bibnamefont
  {Mashkoori}}, \bibinfo {author} {\bibfnamefont {K.}~\bibnamefont
  {Bj\"{o}rnson}}, \ and\ \bibinfo {author} {\bibfnamefont {A.~M.}\
  \bibnamefont {Black-Schaffer}},\ }\bibfield  {title} {\enquote {\bibinfo
  {title} {Impurity bound states in fully gapped d-wave superconductors with
  subdominant order parameters},}\ }\href {\doibase 10.1038/srep44107}
  {\bibfield  {journal} {\bibinfo  {journal} {Sci. Rep.}\ }\textbf {\bibinfo
  {volume} {7}},\ \bibinfo {pages} {44107} (\bibinfo {year}
  {2017})}\BibitemShut {NoStop}%
\bibitem [{\citenamefont {Pershoguba}\ \emph {et~al.}(2015)\citenamefont
  {Pershoguba}, \citenamefont {Bj\"{o}rnson}, \citenamefont {Black-Schaffer},\
  and\ \citenamefont {Balatsky}}]{pershoguba.bjornson.15}%
  \BibitemOpen
  \bibfield  {author} {\bibinfo {author} {\bibfnamefont {S.~S.}\ \bibnamefont
  {Pershoguba}}, \bibinfo {author} {\bibfnamefont {K.}~\bibnamefont
  {Bj\"{o}rnson}}, \bibinfo {author} {\bibfnamefont {A.~M.}\ \bibnamefont
  {Black-Schaffer}}, \ and\ \bibinfo {author} {\bibfnamefont {A.~V.}\
  \bibnamefont {Balatsky}},\ }\bibfield  {title} {\enquote {\bibinfo {title}
  {Currents induced by magnetic impurities in superconductors with spin-orbit
  coupling},}\ }\href {\doibase 10.1103/PhysRevLett.115.116602} {\bibfield
  {journal} {\bibinfo  {journal} {Phys. Rev. Lett.}\ }\textbf {\bibinfo
  {volume} {115}},\ \bibinfo {pages} {116602} (\bibinfo {year}
  {2015})}\BibitemShut {NoStop}%
\bibitem [{\citenamefont {G{\l}odzik}\ and\ \citenamefont
  {Ptok}(2017)}]{glodzik.ptok.17}%
  \BibitemOpen
  \bibfield  {author} {\bibinfo {author} {\bibfnamefont {Sz.}\ \bibnamefont
  {G{\l}odzik}}\ and\ \bibinfo {author} {\bibfnamefont {A.}~\bibnamefont
  {Ptok}},\ }\bibfield  {title} {\enquote {\bibinfo {title} {Quantum phase
  transition induced by magnetic impurity},}\ }\href {\doibase
  10.1007/s10948-017-4360-6} {\bibfield  {journal} {\bibinfo  {journal} {J.
  Supercond. Nov. Magn.}\ } (\bibinfo {year} {2017}),\
  10.1007/s10948-017-4360-6}\BibitemShut {NoStop}%
\bibitem [{\citenamefont {Morr}\ and\ \citenamefont
  {Stavropoulos}(2003)}]{morr.stavropoulos.03}%
  \BibitemOpen
  \bibfield  {author} {\bibinfo {author} {\bibfnamefont {D.~K.}\ \bibnamefont
  {Morr}}\ and\ \bibinfo {author} {\bibfnamefont {N.~A.}\ \bibnamefont
  {Stavropoulos}},\ }\bibfield  {title} {\enquote {\bibinfo {title} {Quantum
  interference between impurities: Creating novel many-body states in s-wave
  superconductors},}\ }\href {\doibase 10.1103/PhysRevB.67.020502} {\bibfield
  {journal} {\bibinfo  {journal} {Phys. Rev. B}\ }\textbf {\bibinfo {volume}
  {67}},\ \bibinfo {pages} {020502} (\bibinfo {year} {2003})}\BibitemShut
  {NoStop}%
\bibitem [{\citenamefont {Morr}\ and\ \citenamefont
  {Yoon}(2006)}]{morr.yoon.06}%
  \BibitemOpen
  \bibfield  {author} {\bibinfo {author} {\bibfnamefont {D.~K.}\ \bibnamefont
  {Morr}}\ and\ \bibinfo {author} {\bibfnamefont {J.}~\bibnamefont {Yoon}},\
  }\bibfield  {title} {\enquote {\bibinfo {title} {Impurities, quantum
  interference, and quantum phase transitions in $s$-wave superconductors},}\
  }\href {\doibase 10.1103/PhysRevB.73.224511} {\bibfield  {journal} {\bibinfo
  {journal} {Phys. Rev. B}\ }\textbf {\bibinfo {volume} {73}},\ \bibinfo
  {pages} {224511} (\bibinfo {year} {2006})}\BibitemShut {NoStop}%
\bibitem [{\citenamefont {Kim}\ \emph {et~al.}(2015)\citenamefont {Kim},
  \citenamefont {Zhang}, \citenamefont {Rossi},\ and\ \citenamefont
  {Lutchyn}}]{kim.zhang.15}%
  \BibitemOpen
  \bibfield  {author} {\bibinfo {author} {\bibfnamefont {Y.}~\bibnamefont
  {Kim}}, \bibinfo {author} {\bibfnamefont {J.}~\bibnamefont {Zhang}}, \bibinfo
  {author} {\bibfnamefont {E.}~\bibnamefont {Rossi}}, \ and\ \bibinfo {author}
  {\bibfnamefont {R.~M.}\ \bibnamefont {Lutchyn}},\ }\bibfield  {title}
  {\enquote {\bibinfo {title} {Impurity-induced bound states in superconductors
  with spin-orbit coupling},}\ }\href {\doibase 10.1103/PhysRevLett.114.236804}
  {\bibfield  {journal} {\bibinfo  {journal} {Phys. Rev. Lett.}\ }\textbf
  {\bibinfo {volume} {114}},\ \bibinfo {pages} {236804} (\bibinfo {year}
  {2015})}\BibitemShut {NoStop}%
\bibitem [{\citenamefont {R\"{o}ntynen}\ and\ \citenamefont
  {Ojanen}(2015)}]{rontynen.ojanen.15}%
  \BibitemOpen
  \bibfield  {author} {\bibinfo {author} {\bibfnamefont {J.}~\bibnamefont
  {R\"{o}ntynen}}\ and\ \bibinfo {author} {\bibfnamefont {T.}~\bibnamefont
  {Ojanen}},\ }\bibfield  {title} {\enquote {\bibinfo {title} {Topological
  superconductivity and high {Chern} numbers in {2D} ferromagnetic {Shiba}
  lattices},}\ }\href {\doibase 10.1103/PhysRevLett.114.236803} {\bibfield
  {journal} {\bibinfo  {journal} {Phys. Rev. Lett.}\ }\textbf {\bibinfo
  {volume} {114}},\ \bibinfo {pages} {236803} (\bibinfo {year}
  {2015})}\BibitemShut {NoStop}%
\bibitem [{\citenamefont {Randeria}\ \emph {et~al.}(2016)\citenamefont
  {Randeria}, \citenamefont {Feldman}, \citenamefont {Drozdov},\ and\
  \citenamefont {Yazdani}}]{randeria.feldman.16}%
  \BibitemOpen
  \bibfield  {author} {\bibinfo {author} {\bibfnamefont {M.~T.}\ \bibnamefont
  {Randeria}}, \bibinfo {author} {\bibfnamefont {B.~E.}\ \bibnamefont
  {Feldman}}, \bibinfo {author} {\bibfnamefont {I.~K.}\ \bibnamefont
  {Drozdov}}, \ and\ \bibinfo {author} {\bibfnamefont {A.}~\bibnamefont
  {Yazdani}},\ }\bibfield  {title} {\enquote {\bibinfo {title} {Scanning
  {Josephson} spectroscopy on the atomic scale},}\ }\href {\doibase
  10.1103/PhysRevB.93.161115} {\bibfield  {journal} {\bibinfo  {journal} {Phys.
  Rev. B}\ }\textbf {\bibinfo {volume} {93}},\ \bibinfo {pages} {161115}
  (\bibinfo {year} {2016})}\BibitemShut {NoStop}%
\bibitem [{\citenamefont {Kim}\ \emph {et~al.}(2017)\citenamefont {Kim},
  \citenamefont {Yoshida}, \citenamefont {Lee}, \citenamefont {Chang},
  \citenamefont {Jeng}, \citenamefont {Lin}, \citenamefont {Haga},
  \citenamefont {Fisk},\ and\ \citenamefont {Hasegawa}}]{kim.yoshida.17}%
  \BibitemOpen
  \bibfield  {author} {\bibinfo {author} {\bibfnamefont {H.}~\bibnamefont
  {Kim}}, \bibinfo {author} {\bibfnamefont {Y.}~\bibnamefont {Yoshida}},
  \bibinfo {author} {\bibfnamefont {Ch.-Ch.}\ \bibnamefont {Lee}}, \bibinfo
  {author} {\bibfnamefont {T.-R.}\ \bibnamefont {Chang}}, \bibinfo {author}
  {\bibfnamefont {H.-T.}\ \bibnamefont {Jeng}}, \bibinfo {author}
  {\bibfnamefont {H.}~\bibnamefont {Lin}}, \bibinfo {author} {\bibfnamefont
  {Y.}~\bibnamefont {Haga}}, \bibinfo {author} {\bibfnamefont {Z.}~\bibnamefont
  {Fisk}}, \ and\ \bibinfo {author} {\bibfnamefont {Y.}~\bibnamefont
  {Hasegawa}},\ }\href@noop {} {\enquote {\bibinfo {title} {Atomic-scale
  visualization of surface-assisted orbital order},}\ } (\bibinfo {year}
  {2017}),\ \Eprint {http://arxiv.org/abs/arXiv:1706.09753} {arXiv:1706.09753}
  \BibitemShut {NoStop}%
\bibitem [{\citenamefont {Reecht}\ \emph {et~al.}(2017)\citenamefont {Reecht},
  \citenamefont {Heinrich}, \citenamefont {Bulou}, \citenamefont {Scheurer},
  \citenamefont {Limot},\ and\ \citenamefont {Schull}}]{reecht.heinrich.17}%
  \BibitemOpen
  \bibfield  {author} {\bibinfo {author} {\bibfnamefont {G.}~\bibnamefont
  {Reecht}}, \bibinfo {author} {\bibfnamefont {B.}~\bibnamefont {Heinrich}},
  \bibinfo {author} {\bibfnamefont {H.}~\bibnamefont {Bulou}}, \bibinfo
  {author} {\bibfnamefont {F.}~\bibnamefont {Scheurer}}, \bibinfo {author}
  {\bibfnamefont {L.}~\bibnamefont {Limot}}, \ and\ \bibinfo {author}
  {\bibfnamefont {G.}~\bibnamefont {Schull}},\ }\href@noop {} {\enquote
  {\bibinfo {title} {Imaging isodensity contours of molecular states with
  {STM}},}\ } (\bibinfo {year} {2017}),\ \Eprint
  {http://arxiv.org/abs/arXiv:1703.05622} {arXiv:1703.05622} \BibitemShut
  {NoStop}%
\bibitem [{\citenamefont {Weismann}\ \emph {et~al.}(2009)\citenamefont
  {Weismann}, \citenamefont {Wenderoth}, \citenamefont {Lounis}, \citenamefont
  {Quaas}, \citenamefont {Ulbrich}, \citenamefont {Dederichs},\ and\
  \citenamefont {Bl\"{u}gel}}]{weismann.wenderoth.09}%
  \BibitemOpen
  \bibfield  {author} {\bibinfo {author} {\bibfnamefont {A.}~\bibnamefont
  {Weismann}}, \bibinfo {author} {\bibfnamefont {M.}~\bibnamefont {Wenderoth}},
  \bibinfo {author} {\bibfnamefont {P.}~\bibnamefont {Lounis}, \bibfnamefont
  {S.and~Zahn}}, \bibinfo {author} {\bibfnamefont {N.}~\bibnamefont {Quaas}},
  \bibinfo {author} {\bibfnamefont {R.~G.}\ \bibnamefont {Ulbrich}}, \bibinfo
  {author} {\bibfnamefont {P.~H.}\ \bibnamefont {Dederichs}}, \ and\ \bibinfo
  {author} {\bibfnamefont {S.}~\bibnamefont {Bl\"{u}gel}},\ }\bibfield  {title}
  {\enquote {\bibinfo {title} {Seeing the {Fermi} surface in real space by
  nanoscale electron focusing},}\ }\href {\doibase 10.1126/science.1168738}
  {\bibfield  {journal} {\bibinfo  {journal} {Science}\ }\textbf {\bibinfo
  {volume} {323}},\ \bibinfo {pages} {1190} (\bibinfo {year}
  {2009})}\BibitemShut {NoStop}%
\bibitem [{\citenamefont {Kawakami}\ and\ \citenamefont
  {Hu}(2015)}]{kawakami.hu.15}%
  \BibitemOpen
  \bibfield  {author} {\bibinfo {author} {\bibfnamefont {T.}~\bibnamefont
  {Kawakami}}\ and\ \bibinfo {author} {\bibfnamefont {X.}~\bibnamefont {Hu}},\
  }\bibfield  {title} {\enquote {\bibinfo {title} {Evolution of density of
  states and a spin-resolved checkerboard-type pattern associated with the
  majorana bound state},}\ }\href {\doibase 10.1103/PhysRevLett.115.177001}
  {\bibfield  {journal} {\bibinfo  {journal} {Phys. Rev. Lett.}\ }\textbf
  {\bibinfo {volume} {115}},\ \bibinfo {pages} {177001} (\bibinfo {year}
  {2015})}\BibitemShut {NoStop}%
\bibitem [{\citenamefont {Jacob}\ \emph {et~al.}(2013)\citenamefont {Jacob},
  \citenamefont {Soriano},\ and\ \citenamefont {Palacios}}]{jacob.soriano.13}%
  \BibitemOpen
  \bibfield  {author} {\bibinfo {author} {\bibfnamefont {D.}~\bibnamefont
  {Jacob}}, \bibinfo {author} {\bibfnamefont {M.}~\bibnamefont {Soriano}}, \
  and\ \bibinfo {author} {\bibfnamefont {J.~J.}\ \bibnamefont {Palacios}},\
  }\bibfield  {title} {\enquote {\bibinfo {title} {Kondo effect and spin
  quenching in high-spin molecules on metal substrates},}\ }\href {\doibase
  10.1103/PhysRevB.88.134417} {\bibfield  {journal} {\bibinfo  {journal} {Phys.
  Rev. B}\ }\textbf {\bibinfo {volume} {88}},\ \bibinfo {pages} {134417}
  (\bibinfo {year} {2013})}\BibitemShut {NoStop}%
\bibitem [{\citenamefont {K\"{u}gel}\ \emph {et~al.}(2015)\citenamefont
  {K\"{u}gel}, \citenamefont {Karolak}, \citenamefont {Kr\"{o}nlein},
  \citenamefont {Senkpiel}, \citenamefont {Hsu}, \citenamefont {Sangiovanni},\
  and\ \citenamefont {Bode}}]{kugel.karolak.15}%
  \BibitemOpen
  \bibfield  {author} {\bibinfo {author} {\bibfnamefont {J.}~\bibnamefont
  {K\"{u}gel}}, \bibinfo {author} {\bibfnamefont {M.}~\bibnamefont {Karolak}},
  \bibinfo {author} {\bibfnamefont {A.}~\bibnamefont {Kr\"{o}nlein}}, \bibinfo
  {author} {\bibfnamefont {J.}~\bibnamefont {Senkpiel}}, \bibinfo {author}
  {\bibfnamefont {P.-J.}\ \bibnamefont {Hsu}}, \bibinfo {author} {\bibfnamefont
  {G.}~\bibnamefont {Sangiovanni}}, \ and\ \bibinfo {author} {\bibfnamefont
  {M.}~\bibnamefont {Bode}},\ }\bibfield  {title} {\enquote {\bibinfo {title}
  {State identification and tunable {Kondo} effect of {MnPc} on {Ag(001)}},}\
  }\href {\doibase 10.1103/PhysRevB.91.235130} {\bibfield  {journal} {\bibinfo
  {journal} {Phys. Rev. B}\ }\textbf {\bibinfo {volume} {91}},\ \bibinfo
  {pages} {235130} (\bibinfo {year} {2015})}\BibitemShut {NoStop}%
\bibitem [{\citenamefont {Island}\ \emph {et~al.}(2017)\citenamefont {Island},
  \citenamefont {Gaudenzi}, \citenamefont {de~Bruijckere}, \citenamefont
  {Burzur\'{\i}}, \citenamefont {Franco}, \citenamefont {Mas-Torrent},
  \citenamefont {Rovira}, \citenamefont {Veciana}, \citenamefont {Klapwijk},
  \citenamefont {Aguado},\ and\ \citenamefont {van~der
  Zant}}]{island.gaudenzi.17}%
  \BibitemOpen
  \bibfield  {author} {\bibinfo {author} {\bibfnamefont {J.~O.}\ \bibnamefont
  {Island}}, \bibinfo {author} {\bibfnamefont {R.}~\bibnamefont {Gaudenzi}},
  \bibinfo {author} {\bibfnamefont {J.}~\bibnamefont {de~Bruijckere}}, \bibinfo
  {author} {\bibfnamefont {E.}~\bibnamefont {Burzur\'{\i}}}, \bibinfo {author}
  {\bibfnamefont {C.}~\bibnamefont {Franco}}, \bibinfo {author} {\bibfnamefont
  {M.}~\bibnamefont {Mas-Torrent}}, \bibinfo {author} {\bibfnamefont
  {C.}~\bibnamefont {Rovira}}, \bibinfo {author} {\bibfnamefont
  {J.}~\bibnamefont {Veciana}}, \bibinfo {author} {\bibfnamefont {T.~M.}\
  \bibnamefont {Klapwijk}}, \bibinfo {author} {\bibfnamefont {R.}~\bibnamefont
  {Aguado}}, \ and\ \bibinfo {author} {\bibfnamefont {H.~S.~J.}\ \bibnamefont
  {van~der Zant}},\ }\bibfield  {title} {\enquote {\bibinfo {title}
  {Proximity-induced {Shiba} states in a molecular junction},}\ }\href
  {\doibase 10.1103/PhysRevLett.118.117001} {\bibfield  {journal} {\bibinfo
  {journal} {Phys. Rev. Lett.}\ }\textbf {\bibinfo {volume} {118}},\ \bibinfo
  {pages} {117001} (\bibinfo {year} {2017})}\BibitemShut {NoStop}%
\bibitem [{\citenamefont {Nakosai}\ \emph {et~al.}(2013)\citenamefont
  {Nakosai}, \citenamefont {Tanaka},\ and\ \citenamefont
  {Nagaosa}}]{nakosai.tanaka.13}%
  \BibitemOpen
  \bibfield  {author} {\bibinfo {author} {\bibfnamefont {S.}~\bibnamefont
  {Nakosai}}, \bibinfo {author} {\bibfnamefont {Y.}~\bibnamefont {Tanaka}}, \
  and\ \bibinfo {author} {\bibfnamefont {N.}~\bibnamefont {Nagaosa}},\
  }\bibfield  {title} {\enquote {\bibinfo {title} {Two-dimensional $p$-wave
  superconducting states with magnetic moments on a conventional $s$-wave
  superconductor},}\ }\href {\doibase 10.1103/PhysRevB.88.180503} {\bibfield
  {journal} {\bibinfo  {journal} {Phys. Rev. B}\ }\textbf {\bibinfo {volume}
  {88}},\ \bibinfo {pages} {180503} (\bibinfo {year} {2013})}\BibitemShut
  {NoStop}%
\bibitem [{\citenamefont {M\'{e}nard}\ \emph {et~al.}(2016)\citenamefont
  {M\'{e}nard}, \citenamefont {Guissart}, \citenamefont {Brun}, \citenamefont
  {Trif}, \citenamefont {Debontridder}, \citenamefont {Leriche}, \citenamefont
  {Demaille}, \citenamefont {Roditchev}, \citenamefont {Simon},\ and\
  \citenamefont {Cren}}]{menard.guissart.16}%
  \BibitemOpen
  \bibfield  {author} {\bibinfo {author} {\bibfnamefont {G.~C.}\ \bibnamefont
  {M\'{e}nard}}, \bibinfo {author} {\bibfnamefont {S.}~\bibnamefont
  {Guissart}}, \bibinfo {author} {\bibfnamefont {Ch.}\ \bibnamefont {Brun}},
  \bibinfo {author} {\bibfnamefont {M.}~\bibnamefont {Trif}}, \bibinfo {author}
  {\bibfnamefont {F.}~\bibnamefont {Debontridder}}, \bibinfo {author}
  {\bibfnamefont {R.~T.}\ \bibnamefont {Leriche}}, \bibinfo {author}
  {\bibfnamefont {D.}~\bibnamefont {Demaille}}, \bibinfo {author}
  {\bibfnamefont {D.}~\bibnamefont {Roditchev}}, \bibinfo {author}
  {\bibfnamefont {P.}~\bibnamefont {Simon}}, \ and\ \bibinfo {author}
  {\bibfnamefont {T.}~\bibnamefont {Cren}},\ }\href@noop {} {\enquote {\bibinfo
  {title} {Two-dimensional topological superconductivity in {Pb/Co/Si(111)}},}\
  } (\bibinfo {year} {2016}),\ \Eprint {http://arxiv.org/abs/arXiv:1607.06353}
  {arXiv:1607.06353} \BibitemShut {NoStop}%
\bibitem [{\citenamefont {Lado}\ and\ \citenamefont
  {Fern\'{a}ndez-Rossier}(2016)}]{lado.fernandezrossier.16}%
  \BibitemOpen
  \bibfield  {author} {\bibinfo {author} {\bibfnamefont {J.~L.}\ \bibnamefont
  {Lado}}\ and\ \bibinfo {author} {\bibfnamefont {J.}~\bibnamefont
  {Fern\'{a}ndez-Rossier}},\ }\bibfield  {title} {\enquote {\bibinfo {title}
  {Unconventional {Yu--Shiba--Rusinov} states in hydrogenated graphene},}\
  }\href {\doibase 10.1088/2053-1583/3/2/025001} {\bibfield  {journal}
  {\bibinfo  {journal} {2D Mater.}\ }\textbf {\bibinfo {volume} {3}},\ \bibinfo
  {pages} {025001} (\bibinfo {year} {2016})}\BibitemShut {NoStop}%
\bibitem [{\citenamefont {Hoffman}\ \emph {et~al.}(2015)\citenamefont
  {Hoffman}, \citenamefont {Klinovaja}, \citenamefont {Meng},\ and\
  \citenamefont {Loss}}]{hoffman.klinovaja.15}%
  \BibitemOpen
  \bibfield  {author} {\bibinfo {author} {\bibfnamefont {S.}~\bibnamefont
  {Hoffman}}, \bibinfo {author} {\bibfnamefont {J.}~\bibnamefont {Klinovaja}},
  \bibinfo {author} {\bibfnamefont {T.}~\bibnamefont {Meng}}, \ and\ \bibinfo
  {author} {\bibfnamefont {D.}~\bibnamefont {Loss}},\ }\bibfield  {title}
  {\enquote {\bibinfo {title} {Impurity-induced quantum phase transitions and
  magnetic order in conventional superconductors: Competition between bound and
  quasiparticle states},}\ }\href {\doibase 10.1103/PhysRevB.92.125422}
  {\bibfield  {journal} {\bibinfo  {journal} {Phys. Rev. B}\ }\textbf {\bibinfo
  {volume} {92}},\ \bibinfo {pages} {125422} (\bibinfo {year}
  {2015})}\BibitemShut {NoStop}%
\bibitem [{\citenamefont {Meng}\ \emph {et~al.}(2015)\citenamefont {Meng},
  \citenamefont {Klinovaja}, \citenamefont {Hoffman}, \citenamefont {Simon},\
  and\ \citenamefont {Loss}}]{meng.klinovaja.15}%
  \BibitemOpen
  \bibfield  {author} {\bibinfo {author} {\bibfnamefont {T.}~\bibnamefont
  {Meng}}, \bibinfo {author} {\bibfnamefont {J.}~\bibnamefont {Klinovaja}},
  \bibinfo {author} {\bibfnamefont {S.}~\bibnamefont {Hoffman}}, \bibinfo
  {author} {\bibfnamefont {P.}~\bibnamefont {Simon}}, \ and\ \bibinfo {author}
  {\bibfnamefont {D.}~\bibnamefont {Loss}},\ }\bibfield  {title} {\enquote
  {\bibinfo {title} {Superconducting gap renormalization around two magnetic
  impurities: From {Shiba} to {Andreev} bound states},}\ }\href {\doibase
  10.1103/PhysRevB.92.064503} {\bibfield  {journal} {\bibinfo  {journal} {Phys.
  Rev. B}\ }\textbf {\bibinfo {volume} {92}},\ \bibinfo {pages} {064503}
  (\bibinfo {year} {2015})}\BibitemShut {NoStop}%
\bibitem [{\citenamefont {\v{Z}itko}(2015)}]{zitko.15}%
  \BibitemOpen
  \bibfield  {author} {\bibinfo {author} {\bibfnamefont {R.}~\bibnamefont
  {\v{Z}itko}},\ }\bibfield  {title} {\enquote {\bibinfo {title} {Numerical
  subgap spectroscopy of double quantum dots coupled to superconductors},}\
  }\href {\doibase 10.1103/PhysRevB.91.165116} {\bibfield  {journal} {\bibinfo
  {journal} {Phys. Rev. B}\ }\textbf {\bibinfo {volume} {91}},\ \bibinfo
  {pages} {165116} (\bibinfo {year} {2015})}\BibitemShut {NoStop}%
\end{thebibliography}%

\end{document}